\documentclass{article}
\usepackage{graphicx}
\usepackage{amssymb, amsmath}
\usepackage{amsfonts}
\usepackage{epstopdf}
\title{How Stands Collapse II}
\author{Philip Pearle\\ 
Hamilton College\\
 Clinton, NY 13323, USA\\
e-mail: ppearle@hamilton.edu}

\begin{document}
\date{}
\maketitle
\abstract {I review ten problems associated with the dynamical wave function collapse program, which were described in the first of these two papers. Five of these, the \textit{interaction, preferred basis, trigger, symmetry} and  \textit{superluminal} problems, were discussed as resolved there.  In this volume in honor of Abner Shimony, I discuss the five remaining problems, \textit{tails, conservation law, experimental, relativity, legitimization}.  Particular emphasis is given to the tails problem, first raised by Abner.  The discussion of  legitimization contains a new argument, that the energy density of the fluctuating field which causes collapse should exert a gravitational force.  This force can be repulsive, since this energy density can be negative.  Speculative illustrations of cosmological implications are offered.}   

\section{Introduction and Recapitulation}

	\textit{\qquad \qquad \qquad  All things in the world come from being.  And being comes}
	
	 \negthinspace\negthinspace{\quad\qquad \qquad \textit{from non-being.} \qquad\qquad \qquad \qquad  The Way of Lao Tzu\\
	 	
	 In 1977, a graduate student at the University of Edinburgh's department of Sociology of Science named Bill Harvey (presently Deputy Director of the Scottish Education Funding Council) was doing his PhD thesis, and wrote to physicists working in the field of foundations of quantum theory, including myself, to ask if he could visit and ask questions.  After my interview, which took place at Hamilton College, Bill, two colleagues and I went out to dinner and, as we drove back, I asked him what his PhD thesis was about.  He said: ``Social deviance." 

	In the first of these papers\cite{HSCI}, hereafter referred to as paper I,  as well as in a previous festschritt for Abner Shimony\cite{AbnerFest}, I presented some personal history, my route to becoming a social deviant.  Closet deviance, shared by a small (but growing, I hope) group of physicists,   
 is the belief that standard quantum theory, handed down on Mount Copenhagen, while a most marvelous set of laws, has conceptual flaws.  Outright deviance is the temerity to try and do something about it.  
 
 	(Parenthetically, Abner Shimony, whom I first met in Wendell Furry's office at Harvard around 40 years ago, has over these years been supportive of my apostasy.  Since Abner is jointly a physicist and philosopher,  he is at most half a deviant, since what is deviant in physics is normal in philosophy.) 

	The flaws are encapsulated in the inadequate answer given by standard quantum theory  to what has been called ``the measurement problem," but which I prefer to call ``the reality problem":
	
	\textit{For a closed system of any kind, given a state vector and the Hamiltonian, specify the evolving realizable states and their probabilities of realization.} 
	
	That is, there is no well-defined procedure within standard quantum theory for, at any time,  plucking out from the state vector the possible states which describe what we see around us. At best, in a restricted set of situations, namely measurement situations by human beings, which are a small subset of the full set of situations in the universe created by nature, one can apply procedures that work FAPP  
( ``For All Practical Purposes," a useful acronym coined by John Bell, in his pungent critique of standard quantum theory\cite{BellUncertainty}).  These procedures require additional, ad hoc (which means ``for this case only") information: \textit{this} is the apparatus, \textit{that} is the environment, etc.  
		
	Paper I described the Continuous Spontaneous Localization (CSL) dynamical wave function collapse theory\cite{MeCSL, GPRCSL}.  It consists of two equations.  A \textit{dynamical equation} describes how the state vector evolves under the joint influence of the Hamiltonian and an operator depending upon an arbitrarily chosen fluctuating scalar field $w(\bf{x}, t)$.  A \textit{probability rule} equation gives the probability that this $w(\bf{x}, t)$ is realized in nature.  Then, the answer given by CSL to the measurement/reality problem is simply:  
	
	\textit{Given any $w(\bf{x}, t)$, a state vector evolving according to the dynamical equation   is a realizable state, and the probability rule gives its probability of realization.} 
	
	  The claim of CSL is ``what you see (in nature) is what you get (from the theory)."   Among other considerations, in this paper it will be argued that this works well.   
		
\subsection{CSL Lite}

\textit{\negthinspace\negthinspace\quad \qquad \qquad Que ser\'a, ser\'a, whatever will be, will be...}
	
	 \qquad\qquad\qquad\quad  Jay Livingstone and Ray Evans, sung by Doris Day\\
		
	In order that this paper be self contained, some of paper I's discussion of CSL will be repeated here,  First comes ``CSL lite,"  a simplified formulation which illustrates essential features.  An initial state vector 
\begin{eqnarray}\label{1}
|\psi, 0\rangle=\sum_{n=1}^{N}c_{n}|a_{n}\rangle  
\end{eqnarray} 	  
(the $|a_{n}\rangle$ are eigenstates of an operator $A$ with nondegenerate eigenvalues $a_{n}$) evolves according to the \textit{dynamical equation}
{\setlength\arraycolsep{2pt}
\begin{eqnarray}\label{2}
|\psi, t\rangle_{w}&\equiv&e^{\frac{1}{4\lambda}\int_{0}^{t}dt'[w(t')-2\lambda A]^{2}}|\psi, 0\rangle\nonumber\\
&=&\sum_{n=1}^{N}c_{n}|a_{n}\rangle e^{-\frac{1}{4\lambda}\int_{0}^{t}dt'[w(t')-2\lambda a_{n}]^{2}}.  
\end{eqnarray}}
 \noindent In Eq. (\ref{2}),  $w(t)$ is a sample random function of white noise type, and $\lambda$ characterizes the collapse rate. The state vector given by (\ref{2}) is not normalized to 1, so one must remember to normalize it when calculating expectation values, the density matrix, etc.
 
 	The probability associated to $|\psi, t\rangle_{w}$ is given by the \textit{probability rule}
 \begin{equation}\label{3}
P_{w}(t)Dw\equiv  _{w}\negthinspace\negthinspace\negthinspace\langle \psi, t|\psi, t\rangle_{w}Dw=\sum_{n=1}^{N}|c_{n}|^{2} e^{-\frac{1}{2\lambda}\int_{0}^{t}dt'[w(t')-2\lambda a_{n}]^{2}}Dw.  
\end{equation}
\noindent To see that the integrated probability is 1, discretize the time integral in Eq. (\ref{3}), so that it appears as a product of gaussians and, using 
\[ Dw\equiv \frac{dw(0)}{\sqrt{2\pi\lambda/\Delta t}}\frac{dw(\Delta t))}{\sqrt{2\pi\lambda/\Delta t}}...\frac{dw(t))}{\sqrt{2\pi\lambda/\Delta t}}, 
\]
\noindent  integrate over all $dw(n\Delta t)$ from $-\infty$ to $\infty$.  

	Here is a proof (not given in paper I, where the result was just cited)  that,  as $t\rightarrow \infty$, Eqs.(\ref{2}),(\ref{3}) describe collapse to one of the eigenstates $|a_{m}\rangle$ with probability $|c_{m}|^{2}$ .  
	
	Consider first the special class of $w(t)$, labeled $w_{a}(t)$, which have the asymptotic behavior
\[\lim_{T\rightarrow \infty}(2\lambda T)^{-1}\int_{0}^{T}dt w_{a}(t)\rightarrow  a, 
\]
\noindent where $a$ is a constant.  Write $w_{a}(t)=w_{0}(t)+2\lambda a$, and define 
\[(2\lambda T)^{-1}\int_{0}^{T}dt w_{0}(t)\equiv\epsilon (T),
\]
so $\lim_{T\rightarrow \infty}\epsilon (T)\rightarrow 0$.   Then Eq.(\ref{3}) may be written
 \begin{equation}\label{4}
P_{w}(t)=\sum_{n=1}^{N}|c_{n}|^{2} e^{-\frac{1}{2\lambda}\int_{0}^{t}dt'w_{0}^{2}(t')}e^{-2\lambda t(a- a_{n})[2\epsilon (t)+(a- a_{n})]}.  
\end{equation}

	 If $a\neq a_{n}$ for any $n$, the probability density (\ref{4}) vanishes for $t\rightarrow \infty$, since it is a sum of terms which  vanish as $\exp- 2\lambda t(a- a_{n})^{2}$.  The (normalized) state vector corresponding to such a $w_{a}(t)$, as given by Eq.(\ref{2}), is generally not a collapsed state, but its asymptotic probability of occurrence is zero.  
	 
	 If $a=a_{m}$, Eqs.(\ref{2}),(\ref{3}) respectively become
{\setlength\arraycolsep{2pt} 	 
\begin{eqnarray}\label{5}
|\psi, t\rangle_{w}&=&e^{-\frac{1}{4\lambda}\int_{0}^{t}dt'w_{0}^{2}(t')}\bigg[c_{m}|a_{m}\rangle+\nonumber\\
&&\qquad\qquad\qquad\qquad\sum_{n\neq m }^{N}c_{n}|a_{n}\rangle 
e^{-\lambda t(a_{m}-a_{n})[2\epsilon (t)+(a_{m}-a_{n})]}\bigg]\nonumber\\
&\rightarrow& e^{-\frac{1}{4\lambda}\int_{0}^{\infty}dt'w_{0}^{2}(t')}c_{m}|a_{m}\rangle
\end{eqnarray}}
{\setlength\arraycolsep{2pt}
\begin{eqnarray}\label{6}
P_{w}(t)&=&e^{-\frac{1}{2\lambda}\int_{0}^{t}dt'w_{0}^{2}(t')}\bigg[|c_{m}|^{2}+\nonumber\\
&&\qquad\qquad\qquad\qquad\sum_{n\neq m }^{N}|c_{n}|^{2} 
e^{-2\lambda t(a_{m}-a_{n})[2\epsilon (t)+(a_{m}-a_{n})]}\bigg]\nonumber\\
&\rightarrow& |c_{m}|^{2}e^{-\frac{1}{2\lambda}\int_{0}^{\infty}dt'w_{0}^{2}(t')}.  
\end{eqnarray}}

\noindent  Eq.(\ref{5}) shows that collapse to $|a_{m}\rangle$ occurs for any $w_{a_{m}}(t)$.  
When Eq.(\ref{6}) is integrated over all possible $w_{a_{m}}(t)$,  (i., e., over all possible $w_{0}(t)$), the total associated probability is $|c_{m}|^{2}$.  

	There are other possibilities for $w(t)$ other than the $w_{a}(t)$, namely the 
cases for which $T^{-1}\int_{0}^{T}dt w(t)$it has no asymptotic limit.  However, since the probability for the $w_{a_{m}}(t)$'s totals to 1, these possibilities have measure 0.  End of proof.

The density matrix constructed from  (\ref{2}), (\ref{3}) is 
\begin{equation}\label{7}
\rho=\int P_{w}(t)Dw\frac{|\psi, t\rangle_{w}\thinspace_{w}\langle \psi,t|}{_{w}\langle \psi,t|\psi,t\rangle_{w}}=\sum_{n, m=1}^{N}c_{n}c_{m}^{*}|a_{n}\rangle\langle a_{m} |e^{-(\lambda t/2)(a_{n}-a_{m})^{2}}.  
\end{equation}
\noindent Thus, the off-diagonal elements decay at a rate determined by the squared differences of eigenvalues.

	For many mutually commuting operators $A_{k}$, and with a possibly time-dependent Hamiltonian $H(t)$ to boot, the evolution (\ref{2}) becomes 
\begin{equation}\label{8}
	|\psi, t\rangle_{w}\equiv {\cal T}e^{-\int_{0}^{t}dt' \{iH(t')+\frac{1}{4\lambda}\sum_{k}[w_{k}(t')-2\lambda A_{k}]^{2}\}}|\psi, 0\rangle,   
\end{equation}
\noindent where ${\cal T}$ is the time-ordering operator.  With $H=0$,  the probability $\sim\negthinspace \negthinspace_{w}\langle \psi,t|\psi,t\rangle_{w}$ is asymptotically non-vanishing only when $w_{k}(t)$ has its asymptotic value equal to $2\lambda$ multiplied by an eigenvalue of $A_{k}$, for each $k$.  The collapse is to the eigenstate labeled by these joint eigenvalues.  

\subsection{CSL}

 	For full-blown CSL,  the index $k$ corresponds to spatial position ${\bf x}$: $w_{k}(t)\rightarrow w({\bf x}, t)$ is considered to be a physical scalar field.  The commuting operators  $A_{k}\rightarrow A({\bf x})$ are taken to be (proportional to) the mass density operator $M({\bf x})$ ``smeared" over a region of length $a$ around $x$.  Thus, the \textit{dynamical equation} is
\begin{equation}\label{9}
	|\psi, t\rangle_{w}\equiv {\cal T}e^{-\int_{0}^{t}dt' \{iH(t')+\frac{1}{4\lambda}\int d {\bf x}[w({\bf x}, t')-2\lambda A({\bf x})]^{2}\}}|\psi, 0\rangle,  
\end{equation}	
\begin{equation}\label{10}
A({\bf x})\equiv\frac{1}{m_{0}(\pi a^{2})^{3/4}}\int  d {\bf z} e^{-\frac{1}{2a^{2}}({\bf x}-{\bf z})^{2}}M({\bf z}).       
\end{equation}
\noindent In Eq.(\ref{9}), $m_{0}$ is taken to be the proton's mass, and the choices $\lambda\approx 10^{-16}$sec$^{-1}$, $a\approx 10^{-5}$cm, the values suggested by Ghirardi, Rimini and Weber for their Spontaneous Localization (SL) theory(\cite{GRW}) are taken, although the present experimental situation allows a good deal of latitude\cite{CollettPearle, Adlerexpt}.  The \textit{probability rule} is, as before, 
\begin{equation}\label{11}
P_{w}(t)Dw=  _{w}\negthinspace\negthinspace\negthinspace\langle \psi, t|\psi, t\rangle_{w}  \prod_{{\bf x},t=0}^{t}\frac{dw({\bf x},t)}{\sqrt{2\pi\lambda/\Delta{\bf x}\Delta t}}.  
\end{equation}

	 Thus, for a state which initially is a superposition of states corresponding to different mass density distributions, ideally (i.e., if one neglects  the Hamiltonian evolution, and waits for an infinite time) one state survives under the CSL dynamics.  The greater the mass density distribution differences between the states, the more rapid is the collapse rate.  When describing the collapse competition between macroscopically distinguishable states, the Hamiltonian evolution can have little effect when it is slow compared to the collapse rate, or when it does not materially affect the mass distribution. 
	 	 
\section{Problem's Progress}

	 Paper I discusses a framework for dynamical collapse models begun in the 70's\cite{pearlevarious, Gisinearly}.  I listed 9 problems which were evident then.  Then, SL came along, a well-defined model of instantaneous collapse, which provides a resolution of 4 problems, but raised  one more.   CSL, which was stimulated by the earlier work and by SL, provides a (somewhat different) resolution of these 5 problems.  The 5 problems and their resolutions are:
	 
	 \textit{Interaction problem}: what should be the interaction which gives rise to collapse?  This is specified in Eqs.(\ref{9}, \ref{10}).  
	 
	  \textit{Preferred basis problem}: what are the states toward which collapse tends?  They are eigenstates of the (smeared) mass density operator (\ref{10}).
	  
	   \textit{Trigger problem}: how can it be ensured that the collapse mechanism is ``off " for microscopically distinguishable states, but ``on" for macroscopically distinguishable states?  This is resolved in CSL, as in SL, by having the collapse always ``on."  In CSL, the collapse rate is slow in the microscopic case because the mass density differences are small, and fast in the macroscopic case because the mass density differences are large.  
	   
	    \textit{Symmetry problem}: how to make the collapse mechanism preserve the exchange symmetry properties of fermionic and bosonic wave functions, which was a problem of SL\cite{SquiresDove}?  This is ensured by the symmetry preserving mass density operator in Eq.(\ref{10}).
	    
	     \textit{Superluminal problem}: how can it be ensured that the collapse dynamics does not allow superluminal communication?  Gisin\cite{Gisinearly} pointed out a necessary condition. It is that the density matrix $\rho(t)$,  evolving from an initial density matrix matrix $\rho(0)$ which can be composed from pure state vectors in various ways, only depend upon $\rho(0)$ and not upon this composition.  It is straightforward to see this is satisfied in CSL, since the density matrix, from Eqs. (\ref{9}), (\ref{11}), is 
{\setlength\arraycolsep{2pt}	     	
\begin{eqnarray}\label{12}
\rho(t)&\equiv& \int Dw P_{w}(t)\frac{|\psi, t\rangle_{w}\ _{w}\langle \psi, t|}{\ _{w}\langle \psi, t|\psi, t\rangle_{w}}\nonumber\\
&=&{\cal T}e^{-\int_{0}^{t}dt' \{iH_{L}(t')-iH_{R}(t')+\frac{\lambda}{2}\int d {\bf x}[A_{L}({\bf x})-A_{R}({\bf x})]^{2}\}}\rho(0) 
\end{eqnarray}}
\noindent (the subscripts $L$ or $R$ mean that the operators are to appear to the left or right of $\rho(0)$, and ${\cal T}$ time-reverse orders operators to the right).  The other necessary ingredient is that the interaction not be long-range.  The gravitational and electrostatic interactions are non-local but not long-range. In a relativistic theory, of course, these interactions are local, transmitted with speed $c$.  In a non-relativistic theory, where particles interact via a non-local potential,  the best one can expect is  the prevention of long-range communication. In CSL, the interaction is via the gaussian-smeared local mass density operator (\ref{10}), so it is non-local, but it is not long-range.  
  
  	  In the remainder of this paper I shall  discuss five problems which remained after the advent of CSL, the \textit{tails, experimental, conservation law, relativity} and \textit{legitimization} problems.  They shall be defined when encountered.  I shall spend most time on the tails problem, because it was first raised by Abner.  
	
\section{Tails Problem}
\textit{\qquad \qquad \qquad  With a little bit, with a little bit, ...}
	
\textit{\quad\qquad \qquad\quad\qquad \qquad \qquad My Fair Lady}, A. J. Lerner and F. Loewe \\

	In November 1980, Abner kindly invited me to stay at his home in Wellesley.  We discussed various aspects of my dynamical collapse program.  In the course of the discussion, Abner expressed the point of view that, in a collapse situation involving macroscopically distinguishable alternatives, one cannot justify saying a definite outcome has occurred if the amplitude of the outcome state is not precisely 1 ( i.e., if  the amplitudes of the rest of the states---the ``tails"---are not precisely zero, no matter how small they are).  Outcomes are observed to occur in a finite time, and the framework for collapse models I had developed allowed different models, ones  where the tails vanish in a finite time or in an infinite time. When I was looking for a physical principle to enable selection of one model over another, I bought Abner's argument and seized upon this to make a choice (\cite{pearlevarious}, 1985).  However, Gisin\cite{Gisinearly} had a better physical principle, avoidance of the superluminal problem. He proposed a model in which the superluminal problem is avoided, but for which the collapse time is infinite.  I showed (\cite{pearlevarious}, 1986) that, generally,  solution of the superluminal problem comes with infinite collapse time.  So, CSL entails the tails problem.  
	
	At a conference in Amherst in June 1990, which was the last time many of us saw John Bell, I remarked in an open session at the end of the conference that I had previously phrased the tails situation in CSL, quite poetically I had thought, as ``a little bit of what might have been is always present with what is," at which point John frowned.  But, I went on, I had learned from him not to say this, for one should not express a new theory in an old theory's language, at which he beamed.  
	
	John died on October 1, 1990.  At a memorial session at the end of that month,  Abner, GianCarlo and I gave talks\cite{philconfAbner, philconfGP} about dynamical collapse, which had been championed by John as a conceptually clear alternative to standard quantum theory.  Abner's talk was entitled ``Desiderata for a Modified Quantum Mechanics."   A number of his desiderata involved the tails issue, raising the question as to whether  CSL is indeed conceptually clear, in particular:
	
	\textit{... it should not permit excessive indefiniteness of the outcome, where ``excessive" is defined by considerations of sensory discrimination ... it does not tolerate ``tails" which are so broad that different parts of the range of the variable can be discriminated by the senses, even if very low probability amplitude is assigned to the tail.}
	
	 A decade ago,  in a festschritt for Abner,  GianCarlo and Tullio Weber\cite{Ghirarditails} and I\cite{Pearletails} gave responses to Abner's position (as did Sohatra Sarkar\cite{Sarkar}, who adopted it) --- see also the lucid paper of Albert and Loewer\cite{AlbertLoewer}.  The problem, in a collapse theory with tails, is to provide a well-defined criterion for the existence of possessed properties of macroscopic variables  which coincides with the evidence of, in Abner's words, ``sensory discrimination." 
	
\subsection{Smeared Mass Density Criterion}
	
		Ghirardi and co-worker's response is based upon the smeared mass density (SMD) whose operator is $A({\bf x})$ (Eq.(\ref{10})).  For a state $|\psi\rangle$, their criterion for the SMD at ${\bf x}$ to have a possessed value (or, in their language, ``accessible" value)  is when the ratio ${\cal R}({\bf x})$ of variance of  $A({\bf x})$ to $\langle\psi |A({\bf x})|\psi\rangle^{2}$ satisfies ${\cal R}({\bf x})<<1$:  then one identifies the possessed value of the SMD with  $\langle\psi |A({\bf x})|\psi\rangle$. 
		
		In measurement situations, because of CSL dynamics, the possessed SMD value criterion very rapidly becomes consistent with our own observations of SMD, for macroscopic objects.  For microscopic objects, e.g., in regions where  only  a few particles are cavorting, the SMD does not have a possessed value but, as Abner stressed, the point of the criterion is to serve to compare the theory with our macroscopic experience. 
		
		However, as Ghirardi et. al.  point out, for a macroscopic object in a superposition of two locations,  after a short time undergoing CSL evolution,  ${\cal R}({\bf x})>>1$ in the region where the object in the tail is located,  so the SMD does not have a possessed value there: one would wish for the value 0.  (This presumes there is no air in the region; when air  at STP is present, the SMD possesses a value in agreement  with experience, the air density.)  Nonetheless, although the criterion fails there, $\langle\psi |A({\bf x})|\psi\rangle<<m_{0}/a^{3}$  in that region, which  is consistent with the experienced value 0.  Another place where the criterion fails is in neither location, where there is no mass density, since ${\cal R}({\bf x})=0/0$
		
		One would like the criterion for the SMD to be possessed to include these cases,  since zero mass density is, in principle, a macroscopic observable.   Although the  authors do not give one, it is easy to obtain:  the SMD  possesses the value $\langle\psi |A({\bf x})|\psi\rangle$ if either ${\cal R}({\bf x})<<1$  OR ${\cal R}({\bf x})>>1$  but  $\langle\psi |A({\bf x})|\psi\rangle<<m_{0}/a^{3}$ OR $\langle\psi |A({\bf x})|\psi\rangle=0$.  There still is an ambiguity as to how small is  $<<1$, which I shall try to make precise later, in the context of my own response to Abner's challenge.  
		
		As I wrote to GianCarlo and Abner, I regard this as an elegant answer to the question: ``What is the \textit{minimum} structure  which will allow one to attribute \textit{macroscopic} reality?"  I addressed, and will address here a different question: ``What is the \textit{maximum} structure which will allow one to attribute reality, \textit{both macroscopic and microscopic}?"  
		
\subsection{Qualified Possessed Value Criterion}

	Rather than reprise my previous argument, I wish to take this opportunity to make it more simple and general. My point of view is that a collapse theory is different from standard quantum theory and, as I said to John Bell in Amherst, therefore requires a new language, conceptual as well as terminological.  
	
	The second sentence of Abner's desideratum quoted above utilizes some words and concepts which, while appropriate for standard quantum theory, are inappropriate for CSL: at the end of this discussion, I shall be more specific.  But, in the first sentence, Abner was absolutely right:  a conceptually sound collapse theory with tails must allow an interpretation which provides no ``indefiniteness of the outcome" and that what is crucial to characterize the definite outcome, are ``considerations of sensory discrimination."  
			
	The new language I propose devolves upon the meaning of the words  \textit{correspond} and \textit{possess} which, to emphasize their importance, I shall irritatingly continue to italicize.  For expository reasons, I shall first review the use of these words in classical and standard quantum physics, before addressing their use in a dynamical collapse theory.  
			 
\subsubsection{Classical Theory Language}
				
		  In classical physics, to a physical state of a system \textit{corresponds}  its  ``mathematical descriptor" (e.g.,  a vector in phase space for a mechanical system) and, \textit{corresponding}  to either, every variable \textit{possesses} a value .
						
		When one is in ignorance about the physical state, then every variable \textit{possesses}, not a value but, rather, a probability distribution of values.  However, these \textit{possessed} entities \textit{correspond} to one's state of ignorance of the physical state of the system, not to the (unknown, but existing) physical state of the system.
		
\subsubsection{Standard Quantum Theory Language} 	
				
		With the advent of quantum phenomena, physicists (especially Bohr) tried to maintain as much classical language as possible.  But something had to give.  What gave is the correspondence of the physical state of the system to the mathematical descriptor, the state vector.
					   
		For a microsystem, the notion of \textit{possessed} value of a variable is preserved by the so-called ``eigenstate-eigenvalue link":  a variable has a \textit{possessed} value only if the operator \textit{corresponding} to the variable has the state vector as an eigenstate, and then the \textit{possessed} value is the eigenvalue.   But, generally, only for very few state vectors can a useful variable can be found which has a  \textit{possessed} value.  Even for a system of modest complexity, for the overwhelming majority of state vectors which describe it,  variables which have \textit{possessed} values are of limited interest, e.g.,  the projection operator on the state itself.  	
					
		For a macrosystem, the precisely applied eigenstate-eigenvalue link does not work.  For example, for such variables as the center of mass position of a meter needle, the location of the ink in a symbol on a computer printout, or the excited state of a radiating pixel on a computer screen, a reasonable state vector \textit{corresponding} to an observed  physical state is not an eigenstate of the \textit{corresponding} operator.  However, if the wave function (the projection of the state vector on a basis vector of the operator)  in some sense has a narrow range, one may try to adopt some criterion to assign a  ``near"  \textit{possessed} value to the variable, a value within the range\cite{philconfGP, AlbertLoewer, Pearletails}.  
													
	 By a preparation or a measurement, i.e., a judicious coupling of a microsystem to a macrosystem, one can force a microsystem to change its physical state to a more desirable one.  Initially, the physical state \textit{corresponds} to microscopic variables which \textit{do not} have \textit{possessed} values and macroscopic variables which \textit{do} have (near) \textit{possessed} values.  Afterwards, the physical state \textit{corresponds} to microsystem and macrosystem variables, which both \textit{do} have \textit{possessed} values or near \textit{possessed} values.  
	 	  
	 The problem, of course, is that the state vector corresponding to the physical state is not produced by the theory.  Schr\"odinger's equation evolves the initial state vector into a state vector where 
neither microscopic nor macroscopic variables have \textit{possessed} values.  One might regard the evolved state vector as a sum (superposition) of state vectors, each of which corresponds to a different possible physical state.  
	 
	Because the evolved state vector is not the descriptor of the state of the evolved physical system, there are various positions taken, within the framework of standard quantum theory to make sense of this situation.  
	 
	 One position is to try to maintain the \textit{correspondence} between the state of the physical system and the state vector by introducing the collapse postulate. To try to select the possible physical states, the collapse results, out of the superposition, there may be pressed into service a  (near) \textit{possessed} value criterion for the macroscopic variables, or  properties of a distinguished part of the physical system, the ``environment,"  may be relied upon.  However, these criteria are ad hoc:  for each different situation they require different knowledge outside the theory.   Sometimes selections made can be quite arbitrary e.g., when the superposition of states is a continuum\cite{Stappcontinuum}.   Indeed, the collapse postulate itself is also ill-defined\cite{PearleAlternative} with regard to when and under what circumstances to apply it.  
	 	 
	 Another position is to regard quantum theory solely as a theory of measurement\cite{PeresFuchs}, and the state vector as a calculational tool.   Thus, Heisenberg considered the state vector of a microsystem to be the repository of ``potentia,"  the capability to describe potential outcomes of future experiments.  Schr\"odinger\cite{Schr}, in discussing this position (with which he was not comfortable), called the state vector which evolves after a measurement the ``expectation catalog," in the sense that it tells one what to expect.  To pluck out the macroscopically distinguishable alternatives from the catalog, again one utilizes the (near) \textit{possessed}  value criterion, informed by the experimental situation.   The ambiguity of when to apply it is of no concern: it is any time after the measurement is completed.  The circumstances of application are limited to experimental situations: although what that means is ill-defined, that is also of no concern to people who take such a pragmatic view of the purpose of quantum theory.  

	Suppose one takes this position, or adopts the ensemble interpretation, the position that it is an ensemble of physical states which \textit{corresponds} to the state vector\cite{Ballentine,PearleAlternative}.  One thus gives up the idea of the \textit{correspondence} of the state of the physical system to the state vector.  If one also believes, as did Bohr,  that standard quantum theory cannot be improved upon, one thereby gives up the possibility of the physical system's state having any kind of mathematical descriptor.   Bell was moved to say that  adoption of this position ``is to betray the great enterprise"\cite{BellUncertainty}.   At the least, it certainly is a great break with classical physics ideas.  
		
	A position which does not make that break is the ``histories'' program\cite{histories}.  Here,  the state of a physical system \textit{corresponds} to a mathematical construct different from the state vector, the so-called ``decoherence functional."  Utilizing standard quantum theory structures,  the hope is to have the decoherence functional correspond to variables which occasionally have \textit{possessed} values.  

	In all these cases,  one is ignorant of the outcome of an experiment.   Thus, just as in classical physics, 
\textit{corresponding} to one's state of  ignorance of the physical state,  a viable variable \textit{possesses} a probability distribution of values.  

	For these positions, how is a tails situation treated?  Suppose a state vector evolving in a measurement situation becomes a superposition of two states whose ratio of amplitudes is enormous.  Suppose also that the values \textit{possessed}  by a macroscopic variable characterizing these states, in Abner's words, ``can be discriminated by the senses even if very low probability is assigned to the tail."  This state vector is interpreted as describing a two-outcome measurement, albeit one outcome is much less likely than the other.  (In the histories scheme, which does not use a state vector, a similar interpretation arises.)  When this situation arises in CSL, one needs a different conclusion, that this state vector describes a one-outcome experiment.  This requires a new language. 
	
\subsubsection{Dynamical Collapse Theory Language}
	
	 CSL retains the classical notion that the physical state of a system \textit{corresponds} to 
the state vector.  \textit{Corresponding} to a random field $w({\bf x}, t)$ whose probability of occurrence (\ref{11}) is non-negligible, the dynamics always evolves a realizable state.  Therefore, one is freed from requiring the  (near) eigenstate-eigenvalue link criterion for the purpose of selecting the realizable states.  I suggest that the eigenstate-eigenvalue link criterion be subsumed by a broader concept.  It must be emphasized that this new conceptual structure is only applicable for a theory which hands you macroscopically sensible realizable states, not superpositions of such states.  
   
	In the new language, corresponding to a quantum state, \textit{every variable possesses a distribution of values}, defined as follows.
	
	If the normalized state is $|\psi\rangle$, consider a variable \textit{corresponding} to the operator 
$B$, with eigenvalues $b$.  Denote the eigenvectors $| b, c\rangle$, where $c$ represents eigenvalues of other operators $C$ which commute with  $B$, all comprising a complete set.  The variable's \textit{possessed} distribution is defined to be $Tr_{c}|\langle  b, c |\psi\rangle|^{2}$ ($Tr_{c}$ represents the trace operation over $C$'s eigenstates).  One may generalize this to say that the set of variables corresponding to the complete set of commuting operators \textit{possesses} a joint distribution $|\langle  b, c |\psi\rangle|^{2}$.  

	What does it mean to say that a variable \textit{possesses} a distribution?  I am never sure what it means to ask what something means\cite{shaggydog}, except that it is a request for more discourse.  

	 I  choose to call this a distribution, \textit{not} a probability distribution, even though it has all the properties of a probability distribution.  This is because, in classical physics, a probability distribution is what \textit{corresponds} to a state of ignorance, and that is not the case here.  What is it a distribution of, if not probability?  Following \cite{Pearletails}, one may give the name ``stuff" to a distribution's numerical magnitude at each value of the variable, as a generalization of Bell's quasi-biblical characterization\cite{BellUncertainty}, "In the beginning,  Schr\"odinger tried to interpret his wavefunction as giving somehow the density of the stuff of which the world was made."  
		
	One is encouraged to think of each variable's stuff distribution as something that is physically real.  
The  notion allows retention of the classical idea that, for a physical state, \textit{every } variable \textit{possesses} an entity}. What is different from classical ideas is that the entity is not a number. One may think of this difference as an important part of what distinguishes the quantum world picture from the classical world picture. 

	But, the distribution notion also differs from standard quantum theory, where one is precluded from thinking of simultaneous values of complementary variables.  In the present view, simultaneously, every variable \textit{possesses} its stuff distribution. Complementarity here means that variables whose operators don't commute do not \textit{possess}  joint distributions, but they do jointly \textit{possess} distributions.   
	 	 	
	Here are a few simple examples.  
	
	If $B$ is the position operator of a particular particle, one may think of the associated position-stuff as representing something real flowing in space.  If  the particle undergoes a two-slit interference experiment, something real is going through both slits and interfering. Likewise, for the particle's momentum operator, real momentum-stuff  also flows in momentum space.  The  ``something real"  can be stuff for any variable represented by an operator function of position and momentum, and all these are \textit{possessed} simultaneously.
	   		
	If B is the operator representing spin in the $\hat{n}$ direction of a spin-1/2 particle, one may think of the $\hat{n}$-spin variable as \textit{possessing} something real, $\hat{n}$-spin-stuff \textit{corresponding} to both values $+\hbar/2$ and $-\hbar/2$, in varying amounts.   Just as in classical physics where a spinning object has a  projection of angular momentum on each direction, and all those values are simultaneously \textit{possessed}, the particle state \textit{corresponds} to variables for all directions, all of whose spin-stuff distributions are simultaneously \textit{possessed}.  
There is one direction, $\hat{m}$, in which the $\hat{m}$-spin-stuff distribution has magnitude 1 at value $+\hbar/2$ and magnitude 0 at value $-\hbar/2$ .  In this case, one can use the language that the $\hat{m}$-spin \textit{possesses} the value $+\hbar/2$.  
		 
\subsubsection{Qualified Possessed Value}
	 
	 A criterion is needed for when it is appropriate to promote a macroscopic  variable's \textit{possessed} stuff distribution to a \textit{possessed} value.  This must be done in order to compare the theory with observation, since observers insist that macroscopic variables  \textit{possess} values.  We shall follow Abner's insightful recourse to ``sensory discrimination,"   as well as take sustenance from a remark in a recent article on the Federal Reserve in The New Yorker\cite{NewYorker}:  ``As social scientists have long recognized, we prefer confident statements of fact to probabilistic statements... ."   Here are two probabilistic considerations. 
	  	 	 	 
	  The first consideration is that an observer's quotation of a \textit{possessed}  value of a macroscopic variable, such as location, velocity, rotation, trajectory, color, brightness,  length,  hardness, ... , is not sufficient. It should contain an error bar.  Such a qualification can readily be supplied, although usually it is not.  Thus, CSL need only present a \textit{possessed} value prediction within the observer's supplied error bar, to favorably compare with the observer's value, 
	  	  	  	  	  	  
	  The second consideration is that, when an observer makes a ``confident  statement" about the \textit{possessed}  value of a macroscopic variable (plus error bar), it needs to be qualified in another way.  If this is to be compared with the theory,  there is the implication that anyone who observes this variable will quote the same value.  This is a prediction, an assertion about the observations of other observers in similar  circumstances, and so it  requires qualification  by providing a measure of the confidence one may give to the assertion or, alternatively, to its falsification.  
	  
	 For example, one might confidently say that all observers will see that lamp is on the table, all observers will see that board's thickness is .75 $\pm$ .01", all observers who toss 100 coins will see them not all come up heads, all observers who  spill  water on the floor will not see it jump back up into the glass, all observers will see that a particular star is in the heavens, all observers of me today can see me tomorrow, etc.  However, each statement is not absolutely sure, and  each should be qualified by giving the probability of its falsification, although sometimes that is not so easy to estimate.      
	  
	  In summary,  a statement about an observed variable, should be characterized by three numbers, a possessed value, the error bar associated with that value, and the probability the statement of value plus error bar is false.  We shall use the latter two numbers in conjunction with a macroscopic variable's stuff distribution, to obtain a criterion for assigning a \textit{possessed} value to the macroscopic variable, for comparison with the first number. 
	  
	From the theory, for the state vector of interest, take the stuff-distribution \textit{possessed} by the macroscopic variable of interest, graphed as stuff versus variable value.  From the observation, take the error bar and slide it along the variable value axis until the maximum amount of  stuff lies within the error bar.  (If the variable has a continuous range of values, and $\Delta$ is the error bar, this condition is simply 
\[
\frac{\partial}{\partial b}\int _{b-\frac{1}{2}\Delta}^{b+\frac{1}{2}\Delta}db' Tr_{c}|\langle  b', c |\psi\rangle|^{2}=Tr_{c}|\langle  b+\frac{1}{2}\Delta, c |\psi\rangle|^{2}-Tr_{c}|\langle  b-\frac{1}{2}\Delta, c |\psi\rangle|^{2}=0.)
\]
\noindent If the amount of stuff outside the error bar is less than the probability of falsification, then the criterion is met, and we shall say that the macroscopic variable has a \textit{qualified possessed value}. 

That value is found, first,  by dividing the variable's distribution  by the amount of stuff within the error bar.  The resulting ``renormalized" distribution is restricted to the error bar range, so that the renormalized amount of stuff within the error bar=1. The \textit{qualified possessed value} is defined as the mean value of the variable calculated with this renormalized  distribution.  This \textit{qualified possessed value} is what is to be compared with the observed \textit{possessed} value, in order to test the validity of the theory.  

 (An alternative is to simply use the variable's unrenormalized distribution to calculate the mean, and call this the variable's \textit{qualified possessed value}, if it lies within the error bar. However, even if the tail amplitude is very small,  the variable's value at the tail could be so large  that it makes a significant contribution to the mean, putting it outside the error bar, which is why this alternative might not produce a \textit{qualified possessed value} which agrees with observation, in circumstances where it ought.)
 
 \subsubsection{Comparison With Observation}
	
	Consider a simple example, a dust particle modeled by a sphere of mass density 1gm/cc and radius $10^{-4}$ cm.  Suppose  the variable of interest is the center of mass position of the sphere.  According to CSL\cite{CollettPearle}, its center of mass wave packet achieves an equilibrium width of $\approx 10^{-8}$cm in about 0.6sec, due to the competition between spreading caused by the Schr\"odinger evolution and  contracting caused by the collapse evolution.  Suppose the dust particle has that equilibrium width.
	
	 Suppose somehow the particle is put into a superposition of two states of equal amplitude, where the centers of mass are further apart than the radius.  According to CSL, the collapse rate  $R\approx\lambda\times$(number of nucleons within a volume $a^{3})\times $(number of nucleons within the sphere).  Thus, $R\approx 10^{-16}\times 6\cdot 10^{8}\times 2.5\cdot 10^{12}=1.5\cdot10^{5}$sec$^{-1}$. Since the tail's squared amplitude $\sim\exp-Rt$, when 
 1 msec has passed, this is $\approx\exp-150$ (and is overwhelmingly likely to be rapidly going down). 
 Therefore, after  1 msec, the typical state describes a center of mass stuff distribution which consists of a packet \textit{corresponding} to squared amplitude $\approx 1$ and width $\approx 10^{-8}$cm at one location and squared amplitude $\approx\exp-150$ and width $\approx 10^{-8}$cm at the other location. 
 	
	 We wish to know whether a \textit{qualified possessed value} of the center of mass exists, according to the criterion and, if so, if it agrees with what an observer would say.  An observer sees the sphere at one location. ``See," is meant literally:  observers use optical light.   We thus can conservatively assign a light  wavelength-restricted error bar of $\approx 10^{-5}$cm.  
	 
	 Moreover, we believe that, in all of human history, all observers in like circumstances would see the same thing.  However, we cannot be absolutely sure of this belief---it hasn't been tested, and can't be.   This suggests that the measure of falsification is not larger than the following stringent estimate.       If all the homo sapiens who have ever lived, an upper estimate of  $\approx10^{11}$ people, were each to spend their whole lives (upper estimate of 100 years $\approx3\cdot10^{12}$msec ) doing nothing else but  observing such a sphere every millisec ($<<$ human perception time of $\approx 100$msec), that in such circumstances only one person once might report seeing something else.  This amounts to a probability of falsification of  $\approx 1/[10^{11}\times3\cdot10^{12}]=3\cdot 10^{-24}\approx\exp-{54}$.  
	 
	 The \textit{qualified possessed value} criterion is met.  The error bar of $10^{-5}$cm. is much larger  than the $10^{-8}$cm spread of the center of mass wave packet.   Essentially all the stuff at the location \textit{corresponding} to squared amplitude $\approx 1$ can be considered to be within the error bar e.g., if the center of mass wave function is $\sim\exp-r^{2}/(10^{-8})^{2}$, this has the value 
 $\exp-10^{6}$ at $r=10^{-5}$cm.  Therefore, the amount of stuff outside the error bar is $\exp-150$, solely due to the tail.  It is much less than the probability of falsification:  $\exp-150<<\exp-{54}$.  Thus, the theory's assignment of \textit{possessed} center of mass location agrees with the observer's assignment.   For a larger object than a mote of dust, it would be satisfied even more easily.   

	 More generally, CSL can be applied to the state of an arbitrarily large fraction of the universe (idealized as isolated), in principle even up to the universe itself.  The physical system should be describable by macroscopic variables with \textit{possessed} values all over space, even if no observer  is there. For the picture given by CSL,  it is helpful\cite{Pearletails} to probe the \textit{corresponding} state vector with  operators representing density variables of every sort:  density of various elementary particle types (i.e., proton, neutron, electron, photon, etc.), density of bound state types, (i.e., nucleii or atoms), density of mass, momentum, velocity, angular momentum, energy,..., integrated over a conveniently sized volume.  For each spatial location of the volume, each such variable will \textit{possess} its distribution.   Because the CSL collapse mechanism rapidly collapses to states where macroscopic objects are well localized, as one moves the probe volume over the space, one  recognizes locations where the variable's distribution exhibits the behavior discussed above, a narrow width packet of total squared amplitude very close to 1, and a small tail.  Thus, one can assign \textit{qualified possessed values} to these variables, and so build up a picture of the macroscopic structure of the system described by the state vector.   
	 
 \subsubsection{Desideratum Revisited}
 
 	I believe the second sentence in Abner's desideratum,
	
	\textit{ ... it does not tolerate ``tails" which are so broad that different parts of the range of the variable can be discriminated by the senses, even if very low probability amplitude is assigned to the tail.},
   
\noindent in referring to the  nature and amplitude of a tail state, uses language appropriate for a quantum theory of measurement, but inappropriate for CSL, which is a quantum theory of reality.  

	Consider the example of a state vector which is a superposition of two macroscopically distinguishable states, a "dominant" state with squared amplitude $1-\epsilon$ and an orthogonal tail state of extremely small squared amplitude $\epsilon$.  According to standard quantum theory, if somehow a measurement of this state can be made in the future (for it is possible in principle, but generally not in practice, to measure a superposition of macroscopically distinguishable states), $\epsilon$ is the probability that the result will correspond to the tail state.  Since repeated measurements do not  \textit{always} yield the dominant state, in a theory where 100\%  reproduceability of measurement results is the criterion for assigning values to variables, one cannot say that the state vector corresponds to the dominant state.   

	In CSL, the tail state and its squared amplitude represent something rather different than a possible outcome of a future measurement and its probability.  The tail state represents an unobservably  small amount of stuff which allows describing the state vector by  (qualified) possessed values assigned to macroscopic variables, consistent with the dominant state. 
 	  	  	
	The role of a tail state's squared amplitude in CSL is best understood by considering the gambler's ruin game analogy to the collapse process.  This was described in paper I but, for completeness, here is a brief recapitulation, in the context of our example.  Two gamblers correspond to the two states. They toss a coin, which corresponds to the fluctuating field.  They exchange money, depending upon the toss outcome, and their net worth fluctuations correspond to fluctuations of the squared  amplitudes.  A result is that a gambler who possesses a fraction $\epsilon$ of the total money has the probability $\epsilon$ of eventually winning all the money.  In particular, even if $\epsilon$ is extremely small, so one of the gamblers has almost lost all his money,  it still is possible that a highly improbable sequence of coin tosses favorable to that gambler can occur, which completely reverses the two gambler's fortunes.  
	
	Analogously, for our example, this means that the dominant state and the tail state have the  probability $\epsilon$ of  spontaneously changing places, what I  call a ``flip."  What does this imply about the picture of nature provided by the theory?  	

	It means that there is a highly improbable possibility that  nature, ``on a whim"  (i.e., by choosing an appropriate field w({\bf x}, t) for a sufficient time interval), can change the universe to a different universe.   In either universe, macroscopic objects have (qualified) possessed values of macroscopic variables.  
	
	Note that such a flip is not triggered by a ``measurement" by anybody:  it is something that can happen spontaneously, at any time.  But, consider a flip, by nature's whim,  occurring right after a measurement with two possible outcomes, where the state vector is as described above.  Before the flip, the universe contains an observer who is sure that result 1 has occurred, and the (qualified) possessed values of macroscopic variables all concur.  After the flip, the universe contains an observer who is sure that result 2 has occurred, and the (qualified) possessed values of macroscopic variables all concur.   
	 
	 To summarize, in the quantum theory of measurement, because one only has the eigenstate-eigenvalue link as a  tool for assigning reality status, one must conclude that a state vector with a tail \textit{cannot} be assigned a reality status consistent with the dominant state.  In CSL, where the dynamics and the (qualified) possessed value criterion are what allows assigning reality status, one concludes that the state vector with a tail \textit{can} be assigned a reality status consistent with the dominant state. There is no problem here, before or after the flip, with assigning a reality status and reconciling an observer's observations with the theory.  
	 
	 Then,  what, in CSL, corresponds to the difficulty faced by the quantum theory of measurement?     The difficulty belongs, not to an observer within the universe, but to some hypothetical being outside the universe (a theoretical physicist?) who keeps track of its state vector.  This being cannot say with 100\% certainty that the realistic universe with a certain history may not at some future time be replaced by another realistic universe with a somewhat different history. Observers within the universe will be oblivious to this (highly improbable) possibility.  And, the theory describes their observations.  
	 
	 Although I have argued here against Abner's position,  I find impressive his insight, a quarter of a century ago, that the tails issue is key to an understanding of important interpretational implications of a dynamical collapse theory.  
	 	 	 	  								
\section{Experimental Problem}	

	CSL is a different theory than standard quantum theory, and so makes different predictions in certain situations.  The problem is to find and perform experiments which test these predictions, with the ultimate goal of either refuting or confirming CSL vis-a-vis standard quantum theory.  
	
	Perhaps the quintessential experimental test involves interference\cite{Pearleinterference,  PearleZeilingerneut}. Suppose an object undergoes a two slit interference experiment.  According to CSL, once there are two spatially separated packets which describe the center of mass  exiting the separated slits, they play the gambler's ruin game and their amplitudes will fluctuate.  Thus, when the packets are brought together once more and their interference is observed, the pattern which results from repeated measurements is predicted to have less contrast (be ``washed out") as compared to the prediction of standard quantum theory.  Indeed, if the packets are separate long enough so that one packet is always dominant, the interference pattern essentially disappears.
	
	 The largest objects so far undergoing interference experiments are $C^{60}$ and $C^{70}$(fullerene or buckyball)\cite{Zeilinger2}. These experiments involved diffraction, so one may visualize a superposition of wave packets emerging from each slit and thereafter all pairs of packets simultaneously compete in the gambler's ruin game.  
The off-diagonal elements of the density matrix between two such packet states decay just as do those for two-slit interference.   The decay factor can be obtained from Eqs.(\ref{10}, \ref{12}) (with the slit size and therefore the packet size less than $a$): it is $\exp-\lambda t n^{2}$, where $n$ is the number of nucleons in the molecule ($n=720$ for $C^{60}$).   The time of flight of a $C^{60}$ was about .05sec and, if one takes the agreement of the observed diffraction pattern with standard quantum theory's prediction to be of 1\% accuracy, this places the limit $\lambda\times.05\times720^{2}<.01$, or 
$\lambda^{-1}>10^{6}$.  A recent proposal\cite{Penrosemirror} to test dynamical collapse, involving the superposition of a mirror in states undisplaced and displaced, has the capability of pushing this limit to $\lambda^{-1}>10^{10}$\cite{Adlermirror}.  Thus, at present, interference experiments have only had a mild impact on CSL.  
				
	The only experiments which, so far, have had an important impact upon CSL,  look for ``spontaneous" increase in particle energy.   It is these experiments which have strongly suggested that a viable CSL must have the mass-density proportional coupling given in Eq.(\ref{10}). 
	 
	Because collapse narrows wave packets, this leads to momentum increase by the uncertainty principle, and therefore energy increase, of all particles.  According to Eqs.(\ref{10}),(\ref{12}),  independently of the potential, the average rate of increase of energy is
\begin{equation}\label{13}
\frac{d\overline{E}_{A}}{dt}=\sum_{k}\frac{3\lambda\alpha_{k}^{2}n_{k}\hbar^{2}}{4m_{k}a^{2}},
\end{equation}
\noindent for any state describing $n_{k}$ particles of type $k$ and mass $m_{k}$ ($\alpha_{k}\equiv(m_{k}/m_{0})$).  However, the SL model, and CSL following it, initially assumed that all particles had the same collapse rate, so that  $\alpha_{k}=1$. 

	More generally, assume that  $\alpha_{p}=1$ for the proton and $\alpha_{k}$  is unknown for other particles.  Eq.(\ref{13})  is an average: CSL predicts that, occasionally, a particle can get a large excitation, which could be detected if a large enough number of particles is observed for a long enough time.
	
 	One can find the probability/sec of a transition from an initial bound state to a final state, from Eq.(\ref{12}) expanded in a series in the size of the bound state divided by $a$.  With the effect of the center of mass wavefunction integrated out, denoting the initial bound state $|\psi_{0}\rangle$ and the final state $|\psi_{f}\rangle$ (bound or free), where these states are eigenstates of the center of mass operator with eigenvalue 0, the transition rate is \cite{PearleSquires}	
\begin{equation}\label{14}
\frac{dP}{dt}=\frac{\lambda}{2a^{2}}|\langle \psi_{f}|\sum_{j,k}\alpha_{k}{\bf r}_{jk}|\psi_{0}\rangle|^{2}+o(\hbox{size}/a)^{4},
\end{equation}		
\noindent where ${\bf r}_{jk}$ is the position operator of the $j$th particle of $k$th type.  	Interestingly, if $\alpha_{k}\sim m_{k}$,  the matrix element of the center of mass operator appears in (\ref{14}), which vanishes.  Then, $dP/dt$ depends upon the much smaller o$(\hbox{size}/a)^{4}$ term. 

	For this reason, experiments which put an upper limit on spontaneous excitation from bound states of atoms or nucleii can constrain the ratios of $\alpha_{k}$'s to be close to the ratios of masses.  
	
	An experiment, which looks for unexplained radiation appearing within a $\approx$1/4kg slab of germanium\cite{Avignone} over a period of about a year, has been applied to a putative CSL ionization of a Ge atom by ejection of a 1s electron\cite{Ge}. Such an excitation should yield a pulse of radiation, 11.1keV from photons emitted by the other electrons in the atom as they cascade down to the new ground state plus the kinetic energy of the ejected electron deposited in the slab.  The probability to ionize the atom is calculated and compared with the experimental upper limit on pulses above 11.1keV.  The result at present is $0\leq \alpha_{e}/\alpha_{N}\leq13m_{e}/m_{N}$, where the subscripts $e$, $N$ refer to the electron and nucleon (proton and neutron parameters are assumed identical). 
	
	In the Sudbury Neutrino Observatory experiment\cite{SNO}, solar neutrinos can collide with deuterium in a sphere 12 meters in diameter. The result is dissociation of deuterium.  (Thereafter, the released neutron, thermalized by collisions, bonds with a deuterium nucleus to form tritium, releasing a 6.25 MeV gamma which then Compton scatters from electrons which emit Cerenkov radiation detected by photodetectors bounding the sphere). The experiment took data for $\approx254$ days, and the observed number of deuterium nucleii was $\approx 5\times 10^{31}$.  The predicted result, using  the standard solar model with neutrino oscillations and the neutrino-deuterium dissociation cross-section, agreed well with the experimental result, within an error range.  Taking this error range as representing an upper limit to CSL excitation of the deuteron, the result is $\alpha_{n}/\alpha_{p}=m_{n}/m_{p}\pm 4\times 10^{-3}$\cite{Snocollapse} (note, $4\times 10^{-3}\approx 3(m_{n}-m_{p})/(m_{n}+m_{p})$).  
	
	These results make plausible  the use of the mass density as the discriminating operator in CSL, $\alpha_{k}= m_{k}/m_{0}$.  The rate of energy increase (\ref{13}) is thus quite small, e.g., over the $13.7\times 10^{9}$yr age of the universe, with the SL values for $\lambda$ and $a$, a single particle acquires energy $E\approx1.3\cdot 10^{-16}m_{k}c^{2}$.
	
		Steve Adler\cite{Adlerexpt} has discussed a number of experiments which could reveal CSL collapse behavior, were $\lambda$ to be substantially larger, than the SL value, say by a factor of $10^{6}$ or more.   I know of only one experimental proposal at present\cite{CollettPearle}, which appears to be currently technically feasible, which could test CSL with the SL value of $\lambda$.  
		
		The idea is that a small sphere will undergo random walk due to CSL\cite{Karolyhazy}.  The expansion of the center of mass wave packet due to Schr\"odinger dynamics is counteracted by the contraction of the wave packet due to CSL dynamics, which results in an equilibrium size for the wave packet. However, since a collapse contraction can occur anywhere within the wave packet, the center of the packet jiggles about.  
		
		Actually, the proposal is rather to observe the random rotation of a small  disc:  the mechanism is similar to that discussed above.   The disc, charged and made of metal, could be suspended and maintained on edge in a Paul trap (an oscillating quadrupole electric field) or, as suggested by Alain Aspect (private communication), a dielectric disc suspended by laser tweezers might    
be feasible.  

	It is a consequence of (\ref{12}) that the ensemble average rms angular deflection of the disc is $\Delta \Theta_{\hbox {CSL}}\approx (\hbar/ma^{2})(\lambda f t^{3}/12)^{1/2}$ ($f$ is a form factor of order 1, depending on the disc dimensions).  For a disc of radius $2\times10^{-5}$cm and thickness $.5\times10^{-5}$cm,  $\Delta \Theta_{\hbox {CSL}}$ diffuses through $2\pi$rad in about 70 sec.  For comparison, according to standard quantum theory,  $\Delta \Theta_{\hbox {QM}}\approx 8\hbar t/\pi mR^{2}$ which,  in 70 sec, is about 100 times less than $\Delta \Theta_{\hbox {CSL}}$.  For example, at an achieved low pressure of $5\times10^{-17}$Torr at liquid helium temperature\cite{Gabrielse}, the mean time between gas molecule collisions with the disc is about 45 minutes, allowing for even a diffusion of the magnitude of $\Delta \Theta_{\hbox{ QM}}$ to be observable.   
			
		I hope that someone  interested in testing fundamental physics will undertake this experiment.  
			
\section{Conservation  Law  Problem}  			
					
	The problem here is that the collapse process appears to violate the conservation laws.  For example,  as discussed in the previous section, particles gain energy from the narrowing of wave functions by collapse.  The resolution is that the conservation laws \textit{are} satisfied when not only the particle contributions, but also the contributions of the the $w({\bf x},t)$ field to the conserved quantities are taken into account.  The easiest way to see this is to take a detour which is interesting in its own right.  
	
	The detour is to discuss a way to quantize the $w({\bf x},t)$ field, and obtain an ordinary Hamiltonian evolution which is mathematically completely equivalent to CSL\cite{PearleIndex}-\cite{PearleCQC}.  For this reason (and because I like the alliteration) I call it a ``Completely Quantized Collapse" (CQC) model although, as will be seen, strictly speaking, this is not what is usually considered as a collapse model.  But, then, it is easy to identify the space-time and rotation generators as conserved quantities, as is usual in a Galilean-invariant quantum theory, and and then extract from them the contributions of the classical $w({\bf x},t)$ field in CSL. 
	
\subsection{CQC}
		
	Define the quantum fields	
{\setlength\arraycolsep{2pt} 
\begin{eqnarray}
W(x)&\equiv&\frac{\lambda^{1/2}}{(2\pi)^{2}}\int d^{4}k[e^{ik\cdot x}b(k)+e^{-ik\cdot x}b^{\dagger}(k)], \label{15}\\
\Pi(x)&\equiv&\frac{i}{2\lambda^{1/2}(2\pi)^{2}}\int d^{4}k[-e^{ik\cdot x}b(k)+e^{-ik\cdot x}b^{\dagger}(k)],\label{16}
\end{eqnarray}}

\noindent 	where $x$ is a four vector, $k\cdot x\equiv \omega t- {\bf k}{\bf\cdot}{\bf x}$ and $[b(k), b^{\dagger}(k')]=\delta^{4}(k-k')$.  It is readily verified that $[W(x), W(x')]=0$,   $[\Pi (x), \Pi(x')]=0$ (the negative energy contribution to these commutators cancels the positive energy contribution) and 
$[W(x), \Pi(x')]=i\delta^{4}(x-x')$.  

	Thus, although $W(x)$ is a quantum field, its value can be simultaneously specified at all space-time events, just like a classical field.  At the space-time event $x$, a basis of eigenstates of  $W(x)$ can be constructed:  $W(x)|w\rangle_{x}=w|w\rangle_{x}$, where $-\infty<w<\infty$.  Using these, a basis 
$|w(x)\rangle\equiv\prod_{x}|w\rangle_{x}$ of eigenstates of the operator $W(x)$ at all events can be constructed, where the eigenstate  $|w(x)\rangle$ can have any eigenvalue at any $x$, and so is labeled by a white noise ``function" $w(x)$.  (For later use, define $|w(x)\rangle_{(a,b)}\equiv\prod_{x, t=a}^{t=b}|w\rangle_{x}$, with  $|w(x)\rangle=|w(x)\rangle_{(-\infty,\infty)}$.)

	If the ``vacuum" state $|0\rangle$ is defined by $b(k)|0\rangle=0$, it follows from (\ref{15}),  (\ref{16}) that
\[
\langle w(x)|\big[W(x)+2i\lambda\Pi(x)\big]|0\rangle= \big[w(x)+2\lambda\frac{\delta}{\delta w(x)}\big]\langle w(x)|0\rangle,
\] 
\noindent so
\begin{equation}\label{17}
\langle w(x)|0\rangle=\exp^{-\frac{1}{4\lambda}\int _{-\infty}^{\infty}d^{4}xw^{2}(x)}, 
\end{equation}
\noindent   with the notation $\int _{a}^{b}d^{4}x\equiv \int _{a}^{b}dt\int _{-\infty}^{\infty}d{\bf x}$.  

	If $|\psi, 0\rangle|0\rangle$  is the initial state, where $|\psi, 0\rangle$ is the initial particle state, define the  evolution in the interaction picture as	 
\begin{equation}\label{18'}
|\Psi,t\rangle\equiv{\cal T}e^{-2i\lambda\int_{0}^{t}d^{4}x'A(x')\Pi(x')}|0\rangle|\psi, 0\rangle
\end{equation} 
\noindent so, from (\ref{17}), (\ref{18'}), 
\begin{equation}\label{19}
 \langle w(x)|\Psi,t\rangle=C(t){\cal T}e^{-\frac{1}{4\lambda}\int_{0}^{t}d^{4}x'[w(x')-2\lambda A(x')]^{2}}|\psi, 0\rangle, 
\end{equation} 
\noindent  where $C(t)=\exp-(4\lambda)^{-1}[\int_{-\infty}^{0}+\int_{t}^{\infty}]d^{4}x'w^{2}(x')]$\label{18} and $A(x)$ is the Heisenberg picture operator,  $A(x)\equiv \exp( iH_{A}t)A({\bf x})\exp(-iH_{A}t)$ ($H_{A}$ is the particle Hamiltonian).  
	
	The expression in Eq.(\ref{19}), apart from the factor $C(t)$, is the CSL interaction picture statevector $|\psi,t\rangle_{w}$, corresponding to the Schr\"odinger picture Eq.(\ref{9}).     Thus it follows from (\ref{19})  that $|\Psi,t\rangle$ may be written as 	
\begin{equation}\label{20}
|\Psi,t\rangle=|\chi\rangle \int Dw_{(0, t)}  |w(x)\rangle_{0,t}|\psi,t\rangle_{w}  
\end{equation}
\noindent  where  $|\chi\rangle\equiv \int Dw_{(-\infty, 0)} Dw_{(t,\infty)}C(t) |w(x)\rangle_{-\infty,0,} |w(x)\rangle_{t,\infty}$ and $Dw_{(0, t)}$ is as defined in Eq.(\ref{11}).  

	Eqs.(\ref{17},\ref{18'},\ref{19}) show that this interaction may be thought of as having the form of a sequence of brief von Neumann measurements.  A ``pointer"  
$w(x)$ is labeled by $x$, and its initial wave function is $\exp-(4\lambda)^{-1}d^{4}xw^{2}(x)$, a very broad gaussian.  The pointers at all $\bf{x}$ with common time $t$ are idle until time $t$, when the brief (duration $dt$)  entanglement interaction occurs (Eq.(\ref{18'}) with the integral over $t$ removed), and they are once again idle.  Each measurement is quite inaccurate, as its variance is $\sim (d^{4}x)^{-1}$.  The resulting wave function Eq.(\ref{19}) describes the state of all pointers having made measurements over the interval $(0,t)$, with $C(t)$ describing the pointers labeled by  $t<\infty$ which will never make measurements, while the pointers labeled by $t>0$ stand waiting to make measurements.  

	I call $|\Psi,t\rangle$, given by Eq.(\ref{20}), the ``ensemble vector."  It is the ``sum" of the (non-orthogonal) CSL states, each multiplied by an eigenstate of the (orthogonal) quantized $w({\bf x},t)$ field.   Therefore, the product states are mutually orthogonal, do not mutually interfere, and they may be unambiguously identified. One may think of the ensemble vector as representing a precisely defined example of Schrodinger's ``expectation catalog," a ``horizontal listing" of the real states of nature, identifiable with the ``vertical listing" of the same states given by CSL. 
	
	The difficulty in making standard quantum theory provide a precise description of the real states of nature, compared with the success of collapse models, was succinctly characterized by John Bell as  ``AND \textit{is not} OR."  But, with CQC,  ``AND \textit{is} OR."  CQC provides a successful model for any interpretation of standard quantum theory, Environmental Decoherence (the $w$-field is the environment), Consistent Histories, Many Worlds, Modal Interpretations, ... .  Key is that, as the particle states  evolve, they are generally not orthogonal, but CQC ``tags" them with eigenstates  $|w(x)\rangle$ which are orthogonal, allowing the eigenstate-eigenvalue link to be successfully employed.  Also  
crucial is that the particle states \textit{can} be regarded as realizable, sensible states of nature, as they are CSL states. 

	A possible benefit of CQC is that it is formulated in standard quantum theory terms, albeit with the strange $W$-field.  This may make it easier  to connect the collapse mechanism with physical mechanisms proposed for other purposes, which are formulated in standard quantum theory terms (see Section 7).  
	
\subsection{Conservation of Energy}

	The free $W(x)$-field time-translation generator is its energy operator:  
	
\begin{equation}\label{21}
H_{w}\equiv\int_{-\infty}^{\infty}d^{4}k\omega b^{\dagger}(k)b(k)=\int_{-\infty}^{\infty}d^{4}x
\dot{W}(x)\Pi(x).
\end{equation}
\noindent (the order of $\dot{W}(x)$ and $\Pi(x)$ can be reversed).  In the Schr\"odinger picture,  the Hamiltonian 
\begin{equation}\label{22}
H=H_{w}+H_{A}+2\lambda  \int d{\bf x}A({\bf x})\Pi({\bf x}, 0)
\end{equation}
\noindent is the time-translation generator, and is conserved.  Because the Hamiltonian is translation and rotation invariant, the momentum and angular momentum operators are likewise conserved (e.g., the momentum operator is $-\int_{-\infty}^{\infty}d^{4}x{\bf \nabla W}(x)\Pi(x)+{\bf P}_{A}$).  Conservation of  energy can be expressed in terms of the constancy of the moment-generating function, 
{\setlength\arraycolsep{2pt} 
\begin{eqnarray}
\langle \Psi, t|e^{-i\beta H}|\Psi, t\rangle&=&\langle \Psi, t|\Psi, t+\beta\rangle=\langle\psi,0|\langle 0|e^{-i\beta H}|0\rangle|\psi,0\rangle\label{23}\\
&=&\langle\psi,0|\langle 0|e^{-i\beta (H_{w}+H_{A})}{\cal T}e^{-i 2\lambda\int_{0}^{\beta}d^{4}xA(x)\Pi(x)}|0\rangle|\psi,0\rangle\nonumber\\
&=&\langle\psi,0|e^{-i\beta H_{A}}{\cal T}e^{-\frac{\lambda}{2}\int_{0}^{|\beta |}d^{4}x
A^{2}(x)}|\psi,0\rangle\nonumber\\
&=&\langle\psi,0| e^{-\big[i\beta H_{A}+|\beta|\frac{\lambda}{2}\int d{\bf x}A^{2}({\bf x})\big]}|\psi,0\rangle.\label{24}
\end{eqnarray}}

\noindent Its  fourier transform, ${\cal P}(E)\equiv (2\pi)^{-1}\int d\beta \exp i\beta E\langle \Psi, t|\exp-i\beta H|\Psi, t\rangle$, is the probability distribution of the energy:
{\setlength\arraycolsep{2pt} 
\begin{eqnarray}\label{25}
{\cal P}(E)&=&\frac{1}{\pi}\langle\psi,0| \frac{1}{(E-H_{A}+i(\lambda/2)\int d{\bf x}A^{2}({\bf x}))}\cdot(\lambda/2)\int d{\bf x}A^{2}({\bf x})\nonumber\\
&&\qquad\qquad\qquad\cdot\frac{1}{(E-H_{A}-i(\lambda/2)\int d{\bf x}A^{2}({\bf x}))}|\psi,0\rangle, 
\end{eqnarray}}

\noindent roughly speaking like the form $\pi^{-1}c/[ (E-\overline{H}_{A})^{2}+c^{2}]$, where $\overline{H}_{A}\equiv \langle\psi,0|H_{A}|\psi,0\rangle$.   In the limit $\lambda\rightarrow 0$, (\ref{25}) reduces to $\delta(E-\overline{H}_{A})$. For $\lambda\neq 0$, the interaction spreads the distribution:   while $\overline{E}=\overline{H}_{A}$, $\overline{E^{2}}=\infty$.  

	Similarly, expressions can be written for the probability distribution of $E_{w}$, $E_{A}$, $E_{I}$ or any sum of two of these, which generally vary with time since these are not constants of the motion.  For example,  it follows from (\ref{19}) that the mean energies are:
{\setlength\arraycolsep{2pt} 
\begin{eqnarray}	
\langle \Psi, t|H_{A}|\Psi, t\rangle&=&\langle\psi,0|{\cal T}_{r}e^{-\frac{\lambda}{2}\int_{0}^{t}dx[A_{L}(x)-A_{R}(x)]^{2}}H_{A}| \psi, 0\rangle,\label{26'}\\
\langle \Psi, t|H_{w}|\Psi, t\rangle&=&\langle\psi,0|\int_{0}^{t}{\cal T}_{r}e^{-\frac{\lambda}{2}\int_{0}^{t'}dx[A_{L}(x)-A_{R}(x)]^{2}}\nonumber\\
&&\qquad\qquad\cdot\int d{\bf x}'[A(x'),[A(x'),H_{A}]| \psi, 0\rangle, \label{27'}\\
\langle \Psi, t|H_{I}|\Psi, t\rangle&=&0\label{28'}.	
\end{eqnarray}}
	
\noindent	where ${\cal T}_{r}$ is the time-reversal ordering operator ($A_{L}$'s are time-reversed, $A_{R}$'s are time-ordered).  Taking the time derivative of (\ref{26'}), (\ref{27'}) shows that $d\overline{H_{A}}/dt=-d\overline{H_{w}}/dt$: in particular, in CSL,  the mean particle kinetic energy increase (\ref{13}) resulting from (\ref{26'})  is compensated by the mean $w$-field energy decrease.  
	
	When collapse has occurred, e.g., following a measurement,  the ensemble vector(\ref{20}) can be written as a sum of macroscopically distinguishable states:  
\begin{equation}\label{26}	
|\Psi, t\rangle=\sum_{n}|\chi\rangle \int_{\Omega_{n}} Dw_{(0, t)}  |w_{n}(x)\rangle_{(0,t)}|\psi,t\rangle_{w_{n}}\equiv \sum_{n}|\Psi, t\rangle_{n},
\end{equation}
\noindent where $|\psi,t\rangle_{w_{n}}$  is a CSL state corresponding to the  $n$th outcome engendered by the field $w_{n}(x)$, and $\Omega_{n}$ is the set of such fields.  Here, not only 
$\thinspace\negthinspace_{(0,t)}\langle w_{m}(x)|w_{n}(x)\rangle_{(0,t)}=0$ for $m\neq n$, but also 
the CSL states are orthogonal (modulo tails), $\thinspace\negthinspace_{w_{m}}\langle \psi,t |\psi,t\rangle_{w_{n}}\approx 0$.  This is because ``macroscopically distinguishable states" means  that the mass density distributions of the CSL states have non-overlapping wave functions (except for tails) in some spatial region(s).  
   
	In this case, energy expressions may be written as the sum of contributions of the separate CSL outcome states.   The product of powers of the energy operators $H_{A}$, $H_{w}$, $H_{I}$, acting on $|\Psi, t\rangle_{n}$,  is essentially orthogonal (that is, up to tails contributions) to states $|\Psi, t\rangle_{m}$, where $m\neq n$.  This is because none of these operators affects the non-overlapping nature of the mass density distribution wave functions. $H_{w}$ doesn't act on $|\psi, t\rangle_{n}$.  $H_{A}$ is the integral over the energy density operator, so it only changes the wave function of the state where the mass density  is non-zero.  $H_{I}$ behaves similarly as it depends upon the integral of the mass density operator.  Thus, for any operator $Q$ formed from these energy operators, 
\[
\langle \Psi, t |Q|\Psi, t\rangle\approx \sum_{n}\thinspace\negthinspace_{n}\langle\Psi, t|Q |\Psi, t\rangle_{n}.  
\]	
\noindent In this manner, generating functions and probabilities can be expressed as the sum of the separate contributions of the CSL states. 

 	Mention should be made of a recent interesting work by Angelo Bassi, Emiliano Ippoliti and Bassano Vacchini\cite{BIV}, who consider a single free particle. The collapse engendering operator is  the position, but modified by adding to it a small term proportional to momentum. The result is that the energy does not increase indefinitely, but reaches an asymptote, in analogy to the behavior of a particle reaching equilibrium with a thermal bath.  The hope is to eventually model the bath and obtain, as discussed here, energy conservation when the particle and bath are both considered.  

\section{Relativity  Problem}  

	The problem is to make a relativistic quantum field theory which describes collapse.  Although a good deal of effort has been expended upon it\cite{Pearleerice}---\cite{NicRim}, there is not a satisfactory theory at present. 
	
		The difficulty is that, while the collapse behavior seems to work just fine, the collapse interaction produces too many particles out of the vacuum, amounting to infinite energy per sec per volume.
		
\subsection{With White Noise}
		
	By  replacing $A({\bf x})$ in Eqs.(\ref {9},\ref {12}) by a  Heisenberg picture quantum field operator $\Phi(x)$ which is a relativistic scalar, replacing $\int_{0}^{t}dt'H(t')$ by a space-time integral over the usual quantum field theory interaction density $V_{I}(x)$ and performing the space-time integral over the region between space-like hypersurfaces $\sigma_{0}, \sigma$, one obtains interaction picture state vector and density matrix evolution equations which are manifestly covariant:			
\begin{equation}\label{27}
	|\psi, \sigma\rangle_{w}\equiv {\cal T}e^{-\int_{\sigma_{0}}^{\sigma}d^{4}x\{iV_{I}(x) +\frac{1}{4\lambda}[w(x)-2\lambda \Phi(x)]^{2}\}}|\psi, \sigma_{0}\rangle,  
\end{equation}			     	
\begin{equation}\label{28}
\rho(\sigma)={\cal T}e^{- \int_{\sigma_{0}}^{\sigma}d^{4}x\{i[V_{IL}(x)-V_{IR}(x)]+ \frac{\lambda}{2}[ \Phi_{L}(x)- \Phi_{R}(x)]^{2}\}}\rho(\sigma_{0}). 
\end{equation}
\noindent The probability density in (\ref {11}) is essentially unchanged, with $t$ replaced by $\sigma$. 
 
	Suppose $\Phi(x)$ is a scalar quantum field.   If  $V_{I}(x)=g\Phi(x):\overline{\Psi}(x)\Psi(x):$, where $\Psi(x)$ is a Dirac fermion quantum field representing some particle type of mass $M$, then the scalar field ``dresses" the particle field, distributing itself around the particle mass density.  Thus, a superposition representing different particle mass distributions will also be a superposition of  different scalar field spatial distributions, and collapse will occur to one or another of these. 
	
	To see what goes wrong, it is easiest to work in what I like to call the the ``collapse interaction picture," where  $\Phi(x)$ is the Heisenberg picture scalar field: this eliminates $V_{I}(x)$'s explicit  presence in Eqs.(\ref {27}, \ref {28}).  In a reference frame where $(\sigma_{0}, \sigma)$ are constant time hyperplanes $(0,t)$, consider the average energy for an  initial density matrix $|\phi\rangle\langle \phi|$:
{\setlength\arraycolsep{2pt} 
\begin{eqnarray}\label{29} 
\overline{H}(t) &=&\hbox {Tr}\big\{H{\cal T}e^{- \frac{\lambda}{2}\int_{\sigma_{0}}^{\sigma}d^{4}x [ \Phi_{L}(x)- \Phi_{R}(x)]^{2}}|\phi\rangle\langle \phi|\big\}\nonumber\\
  &=&\langle \phi| \big\{H-\frac{\lambda}{2}\int_{0}^{t}d^{4}x[\Phi(x)[\Phi(x),H]]+... |\big\}|\phi\rangle\nonumber\\
  &=&\langle \phi|H|\phi\rangle -\frac{i\lambda}{2}\int_{0}^{t}d^{4}x\langle \phi| \big\{[\Phi(x),\dot{\Phi}(x)] |\big\}|\phi\rangle\nonumber\\  
  &=& \langle \phi|H|\phi\rangle+\frac{\lambda}{2}\int_{0}^{t}d^{4}x\delta(0)\nonumber\\
  &=&\langle \phi|H|\phi\rangle+\frac{\lambda t}{2}V\frac{1}{(2\pi)^{3}} \int d{\bf k}.                         
\end{eqnarray}}

\noindent In Eq.(\ref{29}), $\int d{\bf x}=V$ is the volume of space, $\delta(0)=(2\pi)^{-3}\int d{\bf k}\exp i{\bf k}\bf{\cdot}{\bf 0}$ is the sum over modes of the vacuum and is the 0th component of the four-vector 
$(2\pi)^{-3}\int (d{\bf k}/E)(E,{\bf k})$.  Although the energy increase/sec-vol per mode is small, the vacuum gains infinite energy/sec-vol because the vacuum has an infinite number of modes.

	The reason the vacuum is excited can be seen by writing Eq.(\ref{27}) in fourier transform form, mentioned in Eq.(\ref{20}) of paper I: 
{\setlength\arraycolsep{2pt} 
\begin{eqnarray}\label{30} 
|\psi, t\rangle_{w}&=&\int D\eta e^{-\lambda\int_{\sigma_{0}}^{\sigma}d^{4}x \eta^{2}(x)}
e^{i\int_{\sigma_{0}}^{\sigma}d^{4}x\eta(x) w(x)}\nonumber\\
&&\qquad\cdot{\cal T}e^{-i\int_{\sigma_{0}}^{\sigma}d^{4}x\{V_{I}(x)+2\lambda\eta(x)\Phi(x)\}}|\psi, 0\rangle.    	
\end{eqnarray}}

\noindent This can be regarded as an ensemble average over a classical white noise field $\eta(x)$ (the first term in (\ref{30}) is the white noise gaussian probability distribution).  The average is a superposition of unitary evolutions.   The collapse evolution is due to the ``interaction Hamiltonian" density  $\eta(x)\Phi(x)$.  Since $\eta(x)$ is a classical white noise field,  it contains all frequencies and wave numbers in equal amounts.  As a result, because of its interaction with $\Phi(x)$, it excites $\Phi$-particles which possess all possible frequencies and wavelengths out of the vacuum.  

	Indeed, if any mode of the vacuum is excited, for a relativistically invariant theory, all modes must be excited, since that mode looks like another mode in another, equivalent,  reference frame.
	
\subsection{Gaussian Noise}

	To try to remove the vacuum excitation, it is worth considering a noise field  that is not white noise, and therefore doesn't have all frequencies and wavelengths\cite{PearleIndex,PearleDiosi,PearleClinton}. A  generalization of Eqs.(\ref{27},\ref{28}) is
{\setlength\arraycolsep{2pt} 
\begin{eqnarray}
	|\psi, \sigma\rangle_{w}&\equiv&{\cal T}e^{-i\int _{\sigma_{0}}^{\sigma}d^{4}x V_{I}(x)}\nonumber\\	&&\negthinspace\negthinspace\negthinspace\negthinspace\negthinspace\negthinspace\negthinspace\negthinspace
	\cdot e^{-\frac{1}{4\lambda}\int\int_{\sigma_{0}}^{\sigma}d^{4}xd^{4}x' [w(x)-2\lambda \Phi(x)]G(x-x')[w(x')-2\lambda \Phi(x')]}|\psi, \sigma_{0}\rangle,\label{31} \\
\rho(\sigma)&=&{\cal T}e^{- i\int_{\sigma_{0}}^{\sigma}d^{4}x\{[V_{IL}(x)-V_{IR}(x)]} \nonumber\\
&&\negthinspace\negthinspace\negthinspace\negthinspace\negthinspace\negthinspace\negthinspace\negthinspace\cdot e^{- \frac{\lambda}{2}\int\int_{\sigma_{0}}^{\sigma}d^{4}xd^{4}x'  [ \Phi_{L}(x)- \Phi_{R}(x)]G(x-x')[ \Phi_{L}(x')- \Phi_{R}(x')]}\rho(\sigma_{0}), \label{32}
\end{eqnarray}}
\noindent with  
\begin{equation}\label{33}
G(x-x')=\frac{1}{(2\pi)^{4}}\int d^{4}ke^{ik\cdot (x-x')}\tilde{G}(p^{2}),    
\end{equation}
\noindent  where $\tilde{G}(p^{2})\geq0$: if $\tilde{G}(p^{2})=1$, this reduces to the white noise case. 

	CSL, although non-relativistic, can be written in this form.  Put the expression for $A({\bf x})$ from Eq.(\ref{10}) into Eq.(\ref{9}), as well as replace $w({\bf x}, t)$ by  
\[
w({\bf x}, t)\equiv(\pi a^{2})^{-3/4}\int  d {\bf z} e^{-\frac{1}{2a^{2}}({\bf x}-{\bf z})^{2}}w'({\bf z}, t),
\]
\noindent and perform the integral over ${\bf x}$ in the exponent.  The result is 	
{\setlength\arraycolsep{2pt} 
\begin{eqnarray}\label{34}
	|\psi, t\rangle_{w}&\equiv& {\cal T}e^{-i\int_{0}^{t}dt' H(t')}\nonumber\\
&&\negthinspace\negthinspace\negthinspace\negthinspace\negthinspace\negthinspace\negthinspace\negthinspace\negthinspace\negthinspace\negthinspace\negthinspace\negthinspace\negthinspace\negthinspace\negthinspace 
\cdot e^{-\frac{1}{4\lambda}\int\int_{0}^{t} 
dzdz' [w'(z)-\frac{2\lambda }{m_{0}}M(z)]G(z-z')[w'(z')-\frac{2\lambda }{m_{0}}M(z')]}
|\psi, 0\rangle   
\end{eqnarray}}
\noindent where 
\begin{equation}\label{35}
G(z-z')=\delta(t-t')e^{-\frac{1}{4a^{2}}(\bf{z}-\bf{z}')^{2}}.  
\end{equation}	
\subsection{Tachyonic Noise}
	
\subsubsection{$\Phi=$ Free Scalar Field}
	To see how this flexibility can help, reconsider  the calculation of $\overline{H}(t)$ given in (\ref{29}), with the density matrix (\ref{32}), with $(\sigma_{0}, \sigma)$ replaced by $(-T/2,T/2)$ as $T\rightarrow\infty$, and with $\Phi(x)$ a free scalar field of mass $m$ ($V_{I}(x)=0$):  
	{\setlength\arraycolsep{2pt} 
\begin{eqnarray}\label{36} 
\overline{H}(t)
  &=&\langle \phi| \big\{H-\frac{\lambda}{2}\int\int_{-T/2}^{T/2}d^{4}xd^{4}x'G(x-x'){\cal T}_{r}\big\{[\Phi(x)[\Phi(x'),H]]+... \big\}|\phi\rangle\nonumber\\  
  &=&\langle \phi|H|\phi\rangle+\frac{\lambda T}{2}V\tilde{G}(m^{2})\frac{1}{(2\pi)^{3}} \int d{\bf k}                         
\end{eqnarray}}

\noindent (${\cal T}_{r}$ is the time-reversed-ordering operator).  So, if $\tilde{G}(m^{2})=0$, there is no energy creation from the vacuum in this case.  But, then, nothing else happens either!

	This, and further arguments, are most easily understood in terms of Feynman diagrams.  Write the density matrix (\ref{32})  in fourier transform form:
 {\setlength\arraycolsep{2pt} 
\begin{eqnarray}\label{37} 
\rho(\frac{T}{2})&=&\int D\eta e^{-2\lambda\int\int_{-T/2}^{T/2}d^{4}xd^{4}x' \eta(x)G^{-1}(x-x')\eta(x')}\nonumber\\
&&\negthinspace\negthinspace\negthinspace\negthinspace\negthinspace\negthinspace\negthinspace\negthinspace\negthinspace\negthinspace\negthinspace\negthinspace\negthinspace\negthinspace\negthinspace\negthinspace
\negthinspace\negthinspace\negthinspace\negthinspace\negthinspace\negthinspace\negthinspace\negthinspace
\negthinspace\negthinspace\negthinspace\negthinspace\negthinspace\negthinspace\negthinspace\negthinspace
\cdot{\cal T}e^{-i\int_{-T/2}^{T/2}d^{4}x\{V_{I}(x)+\eta(x)2\lambda\Phi(x)\}}\rho(-\frac{T}{2}) 
{\cal T}_{r}e^{i\int_{-T/2}^{T/2}d^{4}x\{V_{I}(x)+\eta(x)2\lambda\Phi(x)\}}.    	
\end{eqnarray}}

\noindent 	The last line of  (\ref{37}) is a unitary transformation, so it can be expanded in a power series, and Wick's theorem used to replace a time-ordered product of operators by a  product of positive and negative frequency normal ordered operators and Feynman propagators.  Then,  $\int D\eta$ can be performed, resulting in  $\int\int_{-T/2}^{T/2}d^{4}xd^{4}x'\eta(x)\eta(x')\rightarrow \int\int_{-T/2}^{T/2}d^{4}xd^{4}x'G(x-x')$:  a term   containing an even number of $\eta(x)$ factors becomes a sum of terms with all possible pairings of 
$\eta(x)$'s replaced by $G$'s. (A term with an odd number of $\eta(x)$ factors vanishes.)  When the integrals over $x$ are performed, the result is the momentum space expression for the sum of Feynman diagrams.  $\tilde{G}(p^{2})$ plays the role of the Feynman propagator for the $\eta$ field. 

	Return to the case of the free $\Phi$ field (i.e., $V_{I}(x)=0$).  Before and after integration over $\eta$, every normal-ordered positive or negative frequency $\Phi$ operator appears in an integral,  
\[\int_{-T/2}^{T/2}d^{4}x\eta(x)\Phi^{\pm}(x)\rightarrow\int_{-T/2}^{T/2}d^{4}xG(x'-x)\Phi^{\pm}(x)=
\tilde{G}(m^{2})\Phi^{\pm}(x')=0,
\] 
\noindent	i.e.,  $G$ and $\Phi$ are orthogonal if $\tilde{G}(m^{2})=0$.  Thus,  the operators disappear  from (\ref{37}).  Then,  $\rho(\frac{T}{2})=C\rho(-\frac{T}{2})$: when the trace is taken, this implies the  c-number $C=1$. 

	It is instructive to look at the  first order in $\lambda$ Feynman graph which describes creation of a $\Phi$-particle from the vacuum, and is responsible for the energy increase given by Eq.(\ref{36}).   Represent the $\Phi$ field by $\wr$ and the $\eta$ propagator by  ${\_ \hspace{-.01in}\_\hspace{-.01in}\_}$.   To lowest order (terms quadratic in  $\eta$), the relevant diagram is $\wr\hspace{-.024in}_{\_ \hspace{-.01in}\_\hspace{-.01in}\_}\hspace{-.035in}\wr$. The $\Phi$ particle created out of the vacuum appears to the left and right sides of the initial density matrix $\rho(-\frac{T}{2})=|0\rangle\langle0|$.  The $\eta$ propagator crosses from one side to the other.  Because the 4-momentum $p$ is conserved (it goes in at the right and out at the left), the diagram is proportional to  $\tilde{G}(p^{2})=\tilde{G}(m^{2})$, with no contribution if  $\tilde{G}(m^{2})=0$.
  
\subsubsection{$\Phi=$ Interacting scalar field}
		 
	With $V_{I}(x)\neq 0$, there can be particle creation out of the vacuum to first order in $\lambda$.  The relevant diagram is $\wr\hspace{-.024in}_{\_ \hspace{-.01in}\_\hspace{-.01in}\_}\hspace{-.035in}\wr$ with a fermion-antifermion pair $\vee$ tacked on to the end of each $\wr$ 
($\wr$ attached at both ends then represents a $\Phi$-particle propagator).  If $p_{1}$ and $p_{2}$ are the outgoing fermion 
4-momenta, the diagram is proportional to $\tilde{G}([(p_{1}+p_{2}]^{2})$.  Vanishing of the contribution of this diagram requires $G$ to vanish for the range of its argument $(2M)^{2}\leq p^{2}<\infty$.  If $M$ can be arbitrarily small, then $\tilde{G}(p^{2})$ \textit{must vanish for all time-like} $p$.  Thus, if we take $\tilde{G}(p^{2})=0$ for $0\leq p^{2}<\infty$, there is no particle creation from the vacuum to first order in $\lambda$:  a space-like 4-momentum (for which  $\tilde{G}(p^{2})$ does not vanish) cannot equal a time-like 4-momentum (of the outgoing fermions).  

	So, the time-like 4-momenta of $\tilde{G}(p^{2})$ are responsible for the energy creation from the vacuum to first order in $\lambda$. In the next subsection we shall see that it is the space-like 4-momenta of $\tilde{G}(p^{2})$ which are responsible for collapse.  
	
	First note that, for diagrams describing collapse, any $\wr$ attached to a ${\_ \hspace{-.01in}\_\hspace{-.01in}\_}$ must be a $\Phi$-particle propagator since, if it represents a free $\Phi$-particle, the diagram's contribution $\sim\tilde{G}(m^{2})=0$.  But, then, this diagram segment's contribution  is 	
\[
\sim \frac{1}{p^{2}-m^{2}+i\epsilon}\tilde{G}(p^{2})=[{\cal P}\frac{1}{p^{2}-m^{2}}-i\pi\delta(p^{2}-m^{2})]\tilde{G}(p^{2})={\cal P}\frac{1}{p^{2}-m^{2}}\tilde{G}(p^{2}).
\]
\noindent Thus, the $\Phi$-particle propagator can be absorbed into the $\eta$ propagator: ${\cal P}[p^{2}-m^{2}]^{-1}\tilde{G}(p^{2})\equiv\tilde{G}'(p^{2})$.  In Feynman diagrams, this means that the $\Phi$-particle propagator line can be replaced by a point: 
for example, the diagram  described in the second sentence of this section,  $\wr\hspace{-.024in}_{\_ \hspace{-.01in}\_\hspace{-.01in}\_}\hspace{-.035in}\wr$ with $\vee$ tacked on to the end of each $\wr$,  
 can be replaced by 
 $\vee_{\hspace{-.05in}\_ \hspace{-.01in}\_\hspace{-.01in}\_}\hspace{-.05in}\vee$ (which, of course, vanishes).
 \subsubsection{$\Phi=$Fermion Density}
 
 	Therefore, we may just consider the model with collapse directly toward fermion density eigenstates, putting $\Phi(x)=:\overline{\Psi}(x)\Psi(x):$ (and setting $V_{I}$ as the usual interaction Hamiltonian for the fermion field with e.g., photons, mesons, ...) into Eqs.(\ref{31}, \ref{32}, \ref{37}).   
	
	In the non-relativistic limit, (\ref{33}) becomes   
 {\setlength\arraycolsep{2pt} 
\begin{eqnarray}\label{38} 
G(x-x')&\rightarrow&\lim_{c\rightarrow\infty}\frac{1}{(2\pi)^{4}}\int dEd{\bf p}e^{iE(t-t')-i{\bf p}\cdot({\bf x}-{\bf x}')}\tilde{G}[\Big(\frac{E}{c}\Big)^{2}-{\bf p}^{2}]\nonumber\\
   &=&\delta(t-t')\frac{1}{(2\pi)^{3}}\int d{\bf p}e^{-i{\bf p}\cdot({\bf x}-{\bf x}')}\tilde{G}(-{\bf p}^{2}).  
\end{eqnarray}}

\noindent With the choice 
\[
\tilde{G}(p^{2})\equiv(4\pi a^{2})^{3/2}\Theta(-p^{2})e^{a^{2} p^{2}}\rightarrow
(4\pi a^{2})^{3/2}e^{-a^{2}{\bf p}^{2}}    
\]	
\noindent ($\Theta$	 is the step function), (\ref{38}) is identical to the CSL form (\ref{35}).  Another interesting  choice is the spectrum $\tilde{G}(p^{2})\equiv \delta (p^{2}+\mu^{2})$ of a tachyon of mass $\mu\approx \hbar/ac\approx 2eV$.  Then, (\ref{38}) becomes $G(x-x')\rightarrow(2\pi)^{-2}\delta(t-t')\sin\mu |x-x'|/|x-x'|/$, which is a perfectly good substitute for the gaussian smearing function.   

	Indeed, with one of these choices, if one regards the non-relativistic limit of 
$:\overline{\Psi}(x)\Psi(x):$ as allowing one to neglect its pair creation and annihilation terms,  the remainder would be the operator $M(x)/m_{0}$, the sum of the number operators for fermion and anti-fermion.  Then  (\ref{32}) would become (\ref{34}): the model would reduce to CSL in the non-relativistic limit.  Unfortunately, one cannot neglect these terms. 

	Alas, in the relativistic model, there is vacuum production of particles to order $\lambda^{2}$.  The expansion of (\ref{37}) to fourth order in $\eta$ produces the vacuum excitation diagram 	
{\large $\vee$}\hspace{-.065in}$_{\_\hspace{-.01in}\_\hspace{-.01in}\_\hspace{-.01in}\_\hspace{-.01in}\_\hspace{-.01in}\_\hspace{-.01in}}\hspace{-.05in}${\large $\vee$}\hspace{-.197in}$^{\_\hspace{-.01in}\_\hspace{-.01in}\_\hspace{-.01in}\_\hspace{-.01in}}$\hspace{.1in}:  two space-like four-momenta of the two $\eta$ propagators can add up to the timelike four-momentum of  the excited fermion pair.

	Thus,  I have given up trying to make a satisfactory relativistic collapse model.  A reason I have gone over this failure in such detail is that it might perhaps stimulate someone to succeed in this endeavor.  Another reason is that, if this failure persists, it helps motivate my fall-back position, the ``Quasi-relativistic" model sketched below\cite{PearleQuasi}.
\subsection{Quasi-relativistic Collapse Model}
	In this model, which has no particle creation from the vacuum,  the state vector and density matrix evolution equations are  Eqs.(\ref{27},\ref{28}), with  {\setlength\arraycolsep{2pt} 
\begin{eqnarray}\label{39} 
\Phi(x)&\equiv&(4\pi a^{2})^{3/4}e^{-\frac{a^{2}}{2}\square}[\overline{\Psi}^{+}(x)\Psi^{-}(x)+\Psi^{+}(x)\overline{\Psi}^{-}(x)]\nonumber\\
&=&\frac{1}{2^{1/2}(\pi a^{2})^{5/4}}\int db^{0}d{\bf b}e^{-\frac{1}{2a^{2}}[(b^{0})^{2}+{\bf b}^{2}]}\nonumber\\
&&\qquad\qquad\cdot[\overline{\Psi}^{+}\Psi^{-}+\Psi^{+}\overline{\Psi}^{-}](t+ib^{0}, {\bf x}+{\bf b}),
\end{eqnarray}}
		
\noindent 	where $\square$ is the D'Alembertian.  This should be compared with the CSL expression for $A$ in Eq.({\ref{10}), written in terms of the particle  annihilation and creation operators $\xi(t,  {\bf x})$,  $\xi^{\dagger}(t,  {\bf x})$:
	
{\setlength\arraycolsep{2pt} 
\begin{eqnarray}\label{40} 
A(x)&\equiv&(4\pi a^{2})^{3/4}e^{-\frac{a^{2}}{2}\nabla^{2}}\xi^{\dagger}(t,  {\bf x})\xi(t,  {\bf x})\nonumber\\
&=&\frac{1}{(\pi a^{2})^{3/4}}\int d{\bf b}e^{-\frac{1}{2a^{2}}{\bf b}^{2}}\xi^{\dagger}\xi(t,  {\bf x}+{\bf b}),
\end{eqnarray}}

\noindent to which (\ref{39}) reduces in the non-relativistic limit (when the anti-particle term $\Psi^{+}\overline{\Psi}^{-}$ is discarded, and the spin degrees of freedom are ignored). Of course, to agree with CSL, when more than one fermion type is considered, there should be a sum of  terms with  coefficients proportional to their masses.  

	The first expression in  (\ref{39}) is manifestly a Lorentz scalar, but the model given by Eqs.(\ref{27},\ref{28},\ref{39}) is not Lorentz invariant.  This is because, while   $\overline{\Psi}\Psi$ does commute with itself at space-like separations,  [$\overline{\Psi}^{+}\Psi^{-}+\Psi^{+}\overline{\Psi}^{-}]$ does not.   Therefore, the time-ordering operation in one Lorentz frame is not the time-ordering operation in another Lorentz frame.  However, it can be shown that, for $-(x-x')^{2}>a^{2}$, the commutator $[\Phi(x),\Phi(x')]\sim \exp-[a/(\hbar/Mc)]$ which for nucleons is $\approx \exp-10^{9}$, i.e., it  ``almost" commutes.  It is in this sense that the model is quasi-relativistic.  
		
	Since there is a preferred reference frame in the model, the one in which time-ordering prevails, it is natural to take it as the co-moving frame in the universe.  Since the earth is not far from the co-moving frame, and the non-relativistic limit of the model is CSL, it so far agrees with experiment.  It would be worthwhile exploring whether there are feasible experiments predicted by the model which would show deviations from relativistic invariance, e.g., experiments with  apparatus moving  rapidly with respect to the preferred frame.  
	
	A number of theoretical proposals\cite{Karolyhazy, Diosi, Penrose, PearleSquiresGrav} have suggested that collapse is related to gravity.  This idea has been buttressed, in the context of CSL, by the experimental evidence for coupling of the fluctuating field $w({\bf x}, t)$ to the mass density operator.  Therefore, there is a positive aspect to a model which is most naturally specialized to the co-moving frame, in that it additionally suggests a cosmological connection for collapse (see the next section).    

	It may also be observed that the relativistic collapse models, which do produce satisfactory collapse behavior in the midst of  unsatisfactory excitation, require the causes of collapse and the space-time locations of the regions where the wave function collapses (rapidly diminishes or grows) to be reference frame dependent\cite{AharonovAlbert, Pearleerice,GGP}. This is not a problem, since the causes and locations of collapse cannot be observed.  But, this differs from the situation in standard quantum field theory, where the amplitudes for particles being in a space-time region do not change when the reference frame changes.   It may then be considered a benefit of this quasi-relativisitic model that it also possesses this same behavior of standard quantum field theory, since the causes and locations associated with collapse are those of the preferred frame.      

\section{Legitimization Problem}

	When, over 35 years ago, as described in paper I, I had the idea of introducing a randomly  fluctuating quantity to cause wave function collapse, I thought, because there are so many things in nature which fluctuate randomly, that when the theory is better developed,  it would  become clear what thing in nature to identify with that randomly fluctuating quantity. Perhaps ironically, this problem of  legitimizing the phenomenological CSL collapse description by tying it in a natural way to  established physics remains almost untouched\cite{Adlerbook}. 
	
	Although, as mentioned in the previous section, various authors, as well as the experimental evidence supporting  coupling of the collapse-inducing fluctuating field to mass density, have suggested a connection between collapse and gravity, it is fair to say that the legitimization problem is still in its infancy. No convincing connection (for example, identification of metric fluctuations, dark matter or dark energy with  $w({\bf x}, t)$) has  yet emerged.  But, I shall give here a new argument that the $w$-field energy density must have a gravitational interaction with ordinary matter, and a perhaps less-convincing argument,  that the  the $w$-field energy density could be cosmologically significant. 
			
\subsection{Gravitational Considerations}
	What happens to the $w$-field energy once it is created, either in small amounts as in measurement situation collapses, or in large amounts as will be suggested below?  Suppose we do not alter the CQC Hamiltonian (\ref{22}).  Then this energy just sits where it was created, and has no other effect on matter.  The picture given in Section 5.1 is that the $w$-field  in an infinitesimal space-time volume is like a pointer making a measurement, which briefly interacts and therefore changes during the measurement, but is unchanged before and after, and its associated energy density has the same behavior.  
	
	But, here is an argument that the CQC Hamiltonian (\ref{22}) must be altered, so that the $w$-field energy  density exerts  a gravitational force on matter. Consider the equation of quasi-classical general relativity,  $G^{\mu,\nu}=-8\pi G \langle \Psi |T^{\mu,\nu}|\Psi\rangle$, i.e.,  $G^{\mu,\nu}$  is classical, but the classical stress tensor is replaced by the quantum expectation value of the stress tensor operator.  Of course, the latter must obey the conservation laws if the equation to to be consistent.  However,  due to the collapse interaction, the expectation value of the particle energy-momentum is not, by itself, conserved.  As discussed in Section 5.2, it is the expectation value of the sum of particle energy-momentum and $w$-field energy-momentum which is conserved.  Therefore,  $T^{\mu,\nu}=T_{A}^{\mu,\nu}+T_{w}^{\mu,\nu}$: the expectation value of the sum of particle and $w$-field stress tensor operators must be utilized.  
	
	In the non-relativistic limit, $G^{0,0}=-8\pi G \langle \Psi |T^{0,0}|\Psi\rangle$ reduces to $\nabla^{2}\phi=4\pi G  \langle \Psi |T^{0,0}|\Psi\rangle$.  Thus, the $w$-field energy density acts just like matter's energy density in creating a gravitational potential, except that the $w$-field energy density can be negative or positive.   
	
	Therefore, when modeling the local behavior of the $w$-field in CQC, and wishing to take into account its gravitational behavior, one ought to modify the CQC Hamiltonian (\ref{22}),  adding a term representing the gravitational interaction of the $w$-field energy density with the matter energy density:
\[H_{G}\equiv -G\int_{-\infty}^{\infty} dx\dot W(x)\Pi(x)\int d{\bf z}\frac{1}{|{\bf x}-{\bf z}|}{\cal H}_{M}({\bf z}).    
\]	
\noindent With this addition, although the $w$-field energy, once created in a volume, still sits in that volume as if nailed in space, it now has an effect on matter, which is repelled/attracted by a region containing  negative/positive  $w$-field energy. 
	
	One could consider further alterations in the local CQC Hamiltonian, to make the $w$-field energy density dynamic, for example, to treat it like a fluid.  Then, it would be gravitationally attracted by matter, or repelled/attracted by itself, if the $w$-field energy density is negative/positive.  One might add a positive constant $w$-field energy to the Hamiltonian, so that the $w$-field energy, although decreased by the collapse interaction, remains positive.  We shall not consider such modifications here.  
	
	To reiterate, the argument here is that compatibility with general relativity requires a gravitational force exerted upon matter by the $w$-field.   
		
\subsection{Cosmological Creation of Negative $w$-Field Energy}
	
	 It was discussed in Section 5.2 that, as the mean energy of matter increases due to collapse, the mean $w$-field energy goes negative by an equal amount.  Thus, if there is an amount of negative $w$-field energy which is of cosmological significance, it would repel matter, and contribute to the observed cosmic acceleration\cite{accel}.   
	  
	 But, as pointed out in Section 4, the mean amount of kinetic energy (\ref{13}) gained by a particle of mass $m$ over the age of the universe is very small, 
\[ E\approx \frac{\lambda a_{0}^{2}}{\lambda_{0}a^{2}}1.3\times10^{-16}mc^{2} 	
\]	 
($\lambda_{0}$, $a_{0}$ are the SL values of $\lambda$ and $a$).   A factor of $10^{16}$ increase in $\lambda/a^{2}$ which makes this energy comparable to $mc^{2}$ would violate already established experimental limits, e.g.,  on ``spontaneous" energy production in atoms or nucleii.  Thus, $w$-field energy created by the collapses accompanying the dynamical evolution of the particles in the universe is not of cosmological significance. 

	However, it is in the spirit of models of the beginning of the universe to imagine that the universe started in a vacuum state, and that it was briefly governed by a Hamiltonian which describes production of particles from the vacuum.  We now illustrate, by a simple model, that negative $w$-field energy of a cosmologically significant amount could be generated in such a scenario. Suppose that, even under such circumstances, the CSL collapse equations apply.  If collapse went on then, as we suppose it does now, the universe would have been in a superposition of the vacuum state and states with various numbers of particles in various configurations, and the collapse mechanism would have been  responsible for choosing the configuration of our present universe.  
	
	This model can also be utilized to describe continuous production of particles as the universe evolves, as in the steady state cosmology.  However, we shall not make that application here.   

	In this simple model, only scalar particles of mass $m$ are produced, and the Hamiltonian is 
\begin{equation}\label{41}
H_{A}=\int_{V} d{\bf x}\{m\xi^{\dagger}({\bf x})\xi({\bf x})+g[\xi({\bf x})+\xi^{\dagger}({\bf x})]\}.
\end{equation}
\noindent where  $\xi({\bf x})$ is the annihilation operator for a scalar particle at ${\bf x}$, $V$ is the volume of the early universe and $g$ is a coupling constant.  With initial state $|\psi,0\rangle=|0\rangle$, with $A(x)=\exp (iH_{A}t)A({\bf x})\exp (-iH_{A}t)$ and $A({\bf x})$  given in Eq.(\ref{10}), we obtain  from  (\ref{26'}), (\ref{41}):
\begin{eqnarray}\label{}
\overline N(t)&\equiv&\langle \Psi,t|\int_{V_{1}}d{\bf x}\xi^{\dagger}({\bf x})\xi({\bf x})|\Psi,t\rangle\nonumber\\
&=&\frac{g^{2}V_{1}}{m^{2}+(\lambda/2)^{2}}
\big\{\lambda t-2(\cos\theta-e^{-\frac{\lambda t}{2}}\cos(\theta+ mt)\big\},\\
\overline Q(t)&\equiv&\langle \Psi,t|\int_{V_{1}}d{\bf x}\frac{1}{2}[\xi^{\dagger}({\bf x})+\xi({\bf x})]|\Psi,t\rangle\nonumber\\
&=&\frac{-gmV_{1}}{m^{2}+(\lambda/2)^{2}}
\big\{1-e^{-\frac{\lambda t}{2}}(\cos mt)+\frac{\lambda}{2m}\sin mt\big\},\\
\overline H_{A}(t)&\equiv&\langle \Psi,t|H_{A}|\Psi,t\rangle=m\overline N(t)+2g\overline Q(t)
\end{eqnarray}
\noindent where $\lambda=\lambda_{0}(m/m_{0})^{2}$,  $V_{1}\subseteq V$, and $\theta\equiv 2\tan^{-1}(2m/\lambda)$.  
One can check that the $\lambda\rightarrow 0$ limit of these equations is the usual oscillatory  quantum mechanical  result (since $\theta=\pi/2$, $m\overline N(t)=-2g\overline Q(t)$, and so $\overline H_{A}(t)=0$.) Also,  all expressions $\rightarrow 0$ as $\lambda\rightarrow \infty$, i.e., in that case the universe remains in the vacuum state due to ``watched pot" or ``Zeno's paradox" behavior (the collapse occurs so fast that there is no chance for the vacuum state to evolve).  

The interesting thing is that the coefficient of the linear increase in $\overline N(t)$ is $\sim g^{2}\lambda$: the Hamiltonian,  acting by itself, generates and annihilates particles,  but without linear growth.   It is the collapse dynamics which, favoring creation over annihilation, is ultimately responsible for creating the matter in the universe, according to this model.  

	Because $d\overline H_{w}/dt= -d\overline H_{A}/dt$, the  mean $w$-field energy $\overline H_{w}(t)$ goes linearly negative.  Moreover, if  ${\cal H}_{w}({\bf x},t)$ and  ${\cal H}_{A}({\bf x},t)$ are the $w$-field and particle energy densities at time $t$, it can be shown that   $d\overline {\cal H}_{w}({\bf x},t)/dt= -d\overline {\cal H}_{A}({\bf x},t)/dt$. Since the initial value of each is zero, we have that $\overline{\cal H}_{w}({\bf x},t)=-\overline{\cal H}_{A}({\bf x},t)$, where the average is over the ensemble of possible universes,  one of which became ours, due to collapse. In each universe, the particle and $w$-field energy densities vary from place to place: in particular, the $w$-field energy density can be negative or positive.  
	
	We can say something interesting about the total  $w$-field and particle energies, $H_{w}$ and  $H_{A}$ in any one universe.  Suppose we divide a particular universe into $N$ equal volumes $\Delta V$, and calculate the mean of the sum $S\equiv$($w$-field energy + particle energy in $k$th volume)/ $\Delta V$ over that universe, i.e., the mean of $S=\sum_{k=1}^{N}(E_{k}/\Delta V)/N=(H_{w}+H_{A})/V$. The probability of $E_{k}$  is independent of $k$, and the mean of  $E_{k}$ is zero, so the mean of $S$ is zero.   By the law of large numbers, as $N\rightarrow\infty$ (which can be achieved by letting $\Delta V $ become infinitesimal), $S$ achieves zero variance. Thus, each universe satisfies  $H_{w}=- H_{A}$. 
	
	This may be thought of as a crude model for the reheating after inflation which produces matter. It should not be taken too seriously: for one thing, one would prefer drawing conclusions from the collapse mechanism applied to the field accompanying an accepted inflationary model.   But, it  suggests that it is possible for the $w$-field energy to be of cosmological significance, that regions of both positive and negative $w$-field energy would then be present, the former attracting matter, the latter repelling matter. If the collapse interaction is not limited to ordinary matter, but includes dark matter, then it suggests that there is a  negative amount of $w$-field energy in the universe equal in magnitude to the mass-energy of all matter.  
\subsection{Some Cosmological Considerations}
	Astronomical observation and theory, which lead to what is called the ``standard model," are woven together in a tight web, so it is rather presumptuous to inject the $w$-field into the mix, especially since the suggestion described at the end of the last subsection is not very detailed.   However, it may stimulate further scrutiny to return to semi-classical gravity, model the quantum expectation values of the matter and $w$-field energy densities in our universe by classical distributions  $\rho_{m}({\bf x},t)$, $\rho_{w}({\bf x},t)$  and discuss a few ways in which the $w$-field might  play a role in affecting the evolution of the universe,  with regard to both fluctuations about the mean behavior and the mean behavior itself. 
	
	With regard to fluctuations, following the suggestion of the model in section 7.2, we suppose, after  the period of particle production in the early universe, that  the $w$-field energy density $\rho_{w}$ is fixed in space,  varies from place to place on the scale of $a$, can be positive or negative, and is initially overlain by mass density $\rho_{m}$.  The negative $w$-field energy density should repel the mass density nearby, the positive  $w$-field energy density should attract it, and so the scenario of matter density fluctuations in the early universe could be affected.  One might speculate that the presently observed voids between galaxies could initially have been sites of negative $w$-field energy density, perhaps initially of scale $a$ which expanded with the universe, that the sites of positive $w$-field energy density could have helped seed initial galactic gravitational collapse and could play a role similar to that of the CDM, etc. If the $w$-field negative energy density (perhaps equal in magnitude to the matter mass-energy, estimated at $\approx \rho_{c}/4$, where $\rho_{c}\equiv 3H_{0}^{2}/8\pi G$ is the critical mass density which makes the universe flat)  is spread fairly uniformly throughout the universe, its gravitational repulsive effect on matter would not seem to have much of an effect on the behavior of formed galaxies, because the density of matter in galaxies is so much greater than $\rho_{c}$. 
		
	As is well known, the  mean behavior of the universe is described by the Friedmann-Robertson-Walker general relativistic  homogeneous isotropic cosmological model, which gives rise to two equations.   One of these can be taken to be the conservation of energy equation relating the universe's scale factor $R$, the energy density $\rho$, and the pressure $p$:  
\begin{equation}\label{44}
\frac{d}{dR}(\rho R^{3})=-3pR^{2}.
\end{equation}	
\noindent  This equation holds for $\rho_{w}$ by itself ($p_{w}=0$) which, glued to space, evolves only due to the expansion of the universe after the period of particle creation has ended, $\rho_{w}\sim R^{-3}(t)$.  Also, $\rho_{m}\sim R^{-3}(t)$ as the matter pressure is negligible.  Let us neglect the radiation density and pressure, and assume a cosmological constant $\rho_{\Lambda}=-p_{\Lambda}$ which also  satisfies (\ref{44}).  
	
	The second equation, the evolution equation for the scale factor $R(t)$, and its current consequence are		
\begin{equation}\label{45}
\frac{\dot R^{2}}{R^{2}}=\frac{8\pi G}{3}\rho-\frac{k}{R^{2}}\equiv H_{0}^{2}\bigg[\frac{\rho}{\rho_{c}}-\frac{k}{R^{2}H_{0}^{2}}\bigg]\Longrightarrow 1=\Omega_{m}+\Omega_{w}+\Omega_{\Lambda}+\Omega_{k}
\end{equation}
\noindent where $R_{0}$ is the present scale factor, $H_{0}\equiv \dot R_{0}/R_{0}$ is Hubble's constant, $k=1,0,-1$ depending respectively upon whether the universe is closed, flat or open, $\Omega_{m} \equiv \rho_{0m}/\rho_{c}$ etc., and  $\Omega_{k} \equiv-k/H_{0}^{2}R_{0}^{2}$.   From (\ref{44}), (\ref{45}) also follows the useful expression
\begin{equation}\label{46}
-\frac{\ddot R}{H_{0}^{2}R}=\frac{\rho}{2\rho_{c}}+\frac{3p}{2\rho_{c}}
\Longrightarrow q_{0}=\frac{1}{2}(\Omega_{m}+\Omega_{w})-\Omega_{\Lambda},   
\end{equation}
\noindent where the deceleration parameter is $q_{0}\equiv -\ddot R_{0}R_{0}/ \dot R_{0}^{2}$.
	
	The matter mass density and $w$-field energy density affect equations (\ref{45}),  (\ref{46}) only through their sum. If we suppose that the $w$-field collapse interaction generates not only the ordinary matter in the universe, but the CDM as well, then $\Omega_{m}+\Omega_{w}=0$.  		
	The consequent result from (\ref{45}),  $\Omega_{\Lambda}+\Omega_{k}=1$, appears to be within 1$\sigma$ of the microwave radiation background data\cite{WMAP}, assuming a flat universe, $\Omega_{k}=0$.  But, when combined with the result from (\ref{46}), $q_{0}=-\Omega_{\Lambda}=-1$, while qualitatively consistent with the observed cosmic acceleration, appears to be 3$\sigma$ from the Hubble plot data\cite{accel}. However, these analyses assume certain prior constraints, and analyzing the prior  $\Omega_{m}+\Omega_{w}=0$ has not received priority.
	
	It is likely that the simple scenario given here will conflict with various astronomical observations and constraints.  There are variants of the model which could be explored to resolve such conflicts, e.g., the parameters $\lambda$, $a$ could vary with time, the $w$-field energy could be made dynamic,  its magnitude could be smaller than  the magnitude of the matter energy (e.g., $\approx 20\%$ of it because the collapse interaction could only occur for ordinary matter and not dark matter), its magnitude could be larger than that of the matter energy (e.g., because collapse could be governed by energy density rather than mass density, and so could occur for light as well as matter), it could play a role in the generation of dark energy, or even be dark energy, etc.  The purpose of this discussion is to illustrate the hope that  progress may be made in legitimizing the phenomenological CSL collapse dynamics by connecting it to the still mysterious contents of the universe.


\begin{thebibliography}{99}
\bibitem{shaggydog}  This brings to mind the ``shaggy dog" story of a man who is driven to find the meaning of life.  After incredible hardship over many years (which takes a long time to recount, but which will be forgone here) he reaches a high mountain in India and obtains an audience with a Guru whom he has been told is the wisest man on earth and who can answer his question. He asks ``Oh sir, what is the meaning of life," and the Guru answers serenely, ``My son, life is a bridge."  At this the man jumps up, visibly upset, and recounts the incredible hardships he has endured (again taking a long time to recount,  but which will be forgone here) ending with, ``And, after all that, what I get is this lousy answer, that life is a bridge?  To which the Guru also jumps up, visibly upset, and says ``Y-y-you mean, life isn't a bridge?"
\end{thebibliography}
\end{document}